\documentclass[twocolumn]{aastex701}

\usepackage{xcolor}

\usepackage[range-phrase=--,range-units=single,list-units=single]{siunitx}
\DeclareSIUnit\bar{bar}
\DeclareSIUnit{\EarthMass}{\ensuremath{\mathit{M}_{\Earth}}}
\DeclareSIUnit{\EarthRadius}{\ensuremath{\mathit{R}_{\Earth}}}

\newcommand{\exorelr}{\mbox{\textsc{ExoReL$^\Re$}}}

\newcommand{\RNUV}{\ensuremath{R_{\mathrm{NUV}}}}
\newcommand{\RVIS}{\ensuremath{R_{\mathrm{VIS}}}}
\newcommand{\RNIR}{\ensuremath{R_{\mathrm{NIR}}}}

\newcommand{\OtwoOtwoCIA}{\ensuremath{\mathrm{O_2}}--\ensuremath{\mathrm{O_2}} CIA}
\newcommand{\Ofour}{\ensuremath{\mathrm{O_4}}}
\newcommand{\HtwoO}{\ensuremath{\mathrm{H_2O}}}
\newcommand{\CHfour}{\ensuremath{\mathrm{CH_4}}}
\newcommand{\COtwo}{\ensuremath{\mathrm{CO_2}}}
\newcommand{\Otwo}{\ensuremath{\mathrm{O_2}}}
\newcommand{\Othree}{\ensuremath{\mathrm{O_3}}}
\newcommand{\Ntwo}{\ensuremath{\mathrm{N_2}}}

\graphicspath{{Figures/}}

\newcommand{\figwidth}{
  \ifdim\columnwidth<\textwidth
    \linewidth
  \else
    0.5\linewidth
  \fi
}

\shorttitle{\Ntwo- vs. \Otwo-dominated atmospheres}
\shortauthors{Burri et al.}

\usepackage{fancyhdr}
\begin{document}

\title{Distinguishing \Ntwo-Dominated from \Otwo-Dominated Atmospheres on Terrestrial Exoplanets with Reflected-Light Spectroscopy}

\author[orcid=0009-0000-3884-5445, gname=Nick, sname=Burri]{Nick Burri}
\affiliation{Jet Propulsion Laboratory, California Institute of Technology, Pasadena, CA 91109, USA}
\affiliation{ETH Zurich, Department of Physics, Otto-Stern-Weg 1, 8093 Zurich, Switzerland}
\email[show]{nick.burri@alumni.ethz.ch}

\author[orcid=0000-0002-1830-8260, gname=Mario, sname=Damiano]{Mario Damiano}
\affiliation{Jet Propulsion Laboratory, California Institute of Technology, Pasadena, CA 91109, USA}
\email[]{mario.damiano@jpl.nasa.gov}

\author[orcid=0000-0003-2215-8485, gname=Renyu, sname=Hu]{Renyu Hu}
\affiliation{Department of Physics and Astronomy, Rutgers, The State University of New Jersey, Piscataway, NJ 08854, USA}
\affiliation{Jet Propulsion Laboratory, California Institute of Technology, Pasadena, CA 91109, USA}
\email[]{rh1071@physics.rutgers.edu}

\begin{abstract}
Reflected-light spectra of Earth-sized exoplanets promise to probe habitability and biosignatures, but interpreting atmospheric oxygen can be challenging because \Otwo\ can arise from both biological and abiotic processes. In addition, reflected-light retrievals exhibit degeneracies among atmospheric composition, clouds, surface reflectance, and gravity. We use the Bayesian retrieval framework \exorelr, leveraging recent model developments that incorporate (i) \Otwo--\Otwo\ collision-induced absorption (CIA), which produces the \Ofour\ bands, and (ii) a wavelength-dependent (step-function) surface albedo parameterization. Focusing on a modern Earth analog, we compare retrievals with and without \OtwoOtwoCIA\ using spectra within \qtyrange{0.25}{1.8}{\micro\meter}, and perform signal-to-noise ratio (SNR) sweeps for two observing configurations motivated by the Habitable Worlds Observatory: a resolved-visible case ($\RVIS = \num{140}$) and a broadband-visible case (three broad bins). Omitting \OtwoOtwoCIA\ allows the retrieval to fit the spectrum with an incorrect \Otwo-dominated atmosphere. Including \OtwoOtwoCIA\ leverages the \Ofour\ bands to break this degeneracy and recover the correct \Ntwo-dominated, \Otwo-rich solution without requiring a tight mass prior. Adopting a robustness criterion of \qty{\geq 95}{\percent} posterior mass in the \Ntwo-dominated region, both configurations require $\mathrm{SNR} \gtrsim \num{20}$ at \qty{0.75}{\micro\meter} ($\mathrm{SNR} \approx \num{14}$ near the \qty{1.27}{\micro\meter} \Ofour\ band for $\RNIR = \num{70}$) to identify the correct dominant gas, with higher cloud fraction increasing this threshold. Reducing near-IR coverage to \qty{1.4}{\micro\meter} prevents recovery of the correct \Ntwo-dominated solution across the explored SNR range, and instead favors a spurious \Otwo-dominated, \COtwo-rich atmosphere. These results support near-IR coverage to \qty{\sim 1.7}{\micro\meter} and $\mathrm{SNR} \gtrsim \num{20}$ at \qty{0.75}{\micro\meter} for robust oxygen interpretation on Earth-like planets.
\end{abstract}


\section{Introduction}\label{sec:introduction}

The search for potentially habitable exoplanets and the characterization of their atmospheres are central goals of modern astrophysics. A key challenge is the detection and interpretation of atmospheric biosignatures: molecular species whose presence may indicate biological activity. To address this challenge, the Astro2020 Decadal Survey prioritized a space-based observatory capable of directly imaging and spectrally characterizing Earth-sized exoplanets in the habitable zones of nearby stars \citep{NASEM_2023}, motivating the Habitable Worlds Observatory (HWO) mission concept.

Reflected-light spectroscopy can constrain atmospheric composition, cloud properties, surface reflectivity, and bulk parameters for directly imaged planets, including non-transiting systems. A growing body of work has demonstrated that Bayesian atmospheric retrieval frameworks can extract meaningful constraints from such spectra and can be used to evaluate observing strategies and instrument trade spaces \citep[e.g.,][]{Lupu_2016,Nayak_2017,Feng_2018,Damiano_2020,Damiano_2022,Robinson_2023,Young_2024,Salvador_2025}. However, in the regime of terrestrial atmospheres, retrievals often face significant degeneracies, as multiple atmospheric, surface, and observational effects can produce similar spectral signatures. In particular, weak constraints on planet mass (and therefore surface gravity) allow trade-offs between column densities, surface pressure, and bulk composition that can qualitatively change the inferred dominant atmospheric constituent \citep{Damiano_2025}. For both modern- and Archean-Earth analogs, \citet{Damiano_2025} showed that when the planet mass is not known to within \qty{\sim 10}{\percent}, retrievals can confuse the dominant and secondary gases (e.g., \Ntwo\ versus \Otwo\ in modern-Earth-like cases, and \Ntwo\ versus \COtwo\ in Archean-like cases).

A prominent example for a biosignature is oxygen. \Otwo\ is often treated as a compelling biosignature because of its link to oxygenic photosynthesis on Earth, yet substantial \Otwo\ can also be produced by abiotic processes under certain conditions \citep[e.g.,][]{Harman_2015,Meadows_2017}. In particular, some abiotic pathways can generate very large \Otwo\ inventories, for example via water photolysis followed by hydrogen escape, potentially leading to \Otwo-dominated atmospheres in which \Otwo\ becomes the primary constituent \citep[e.g.,][]{Wordsworth_2014,Luger_2015}. Throughout this paper, we define an atmosphere as \Otwo-dominated when \Otwo\ accounts for more than half of the atmospheric volume mixing ratio ($\mathrm{VMR}_\Otwo > \num{0.5}$), and as \Ntwo-dominated analogously when $\mathrm{VMR}_\Ntwo > \num{0.5}$. We use ``oxygen-rich'' descriptively for atmospheres with substantial \Otwo, such as the modern Earth case, but without it being the dominant constituent. Because the reflected-light signatures of an \Otwo-dominated atmosphere can resemble those of an inhabited, oxygen-producing world, interpreting an \Otwo\ detection requires separating \Ntwo-dominated, oxygen-rich atmospheres from \Otwo-dominated ones. In reflected light, this distinction can be obscured by degeneracies that allow different atmospheric compositions to fit the same spectrum through correlated adjustments in surface pressure, planetary radius, cloud parameters, surface albedo, and surface gravity when mass constraints are weak \citep[e.g.,][]{Damiano_2025}.

A potentially powerful diagnostic is collision-induced absorption (CIA) from \Otwo--\Otwo\ interactions. In this paper, we refer to the opacity process as \OtwoOtwoCIA\ and to the associated spectral bands as \Ofour. Because \OtwoOtwoCIA\ scales with the square of the \Otwo\ number density, \Ofour\ provides sensitivity to \Otwo\ column density beyond that contained in monomer \Otwo\ bands. \Ofour\ is well characterized in laboratory and theoretical studies \citep[e.g.,][]{Greenblatt_1990,Thalman_2013,Karman_2018}. It has been discussed as a potentially informative feature for terrestrial exoplanet characterization in multiple observing geometries, including reflected light, and it has been explored as a pressure and context diagnostic in transit transmission spectroscopy (including for assessing abiotic oxygen buildup) as well as in mid-infrared studies relevant to thermal-emission characterization \citep[e.g.,][]{Misra_2014,Schwieterman_2016,Fauchez_2020,Alei_2022}. However, the specific impact of \OtwoOtwoCIA\ on Bayesian reflected-light retrieval outcomes for terrestrial planets has not been systematically quantified.

Here we quantify how including \OtwoOtwoCIA\ and varying observing configurations affect our ability to discriminate between \Ntwo-dominated, oxygen-rich atmospheres and \Otwo-dominated atmospheres for a modern Earth analog in reflected light. We show that \OtwoOtwoCIA\ can provide sufficient additional spectral leverage to favor the correct \Ntwo-dominated solution even when planet mass (and thus surface gravity) is not constrained, complementing the mass-precision requirements identified by \citet{Damiano_2025}. We use \exorelr\ \citep{Damiano_2020,Damiano_2022}, extended to include \Ofour\ opacities and a wavelength-dependent surface albedo parameterization relative to previous studies, and we perform retrieval comparisons and signal-to-noise ratio (SNR) sweeps for HWO-motivated spectral configurations.

This paper is organized as follows. Section~\ref{sec:methods} describes the retrieval framework, the model extensions implemented prior to this work, and the atmospheric and observational scenarios considered. Section~\ref{sec:results} presents the retrieval results, focusing on the role of \Ofour\ and the dependence of compositional constraints on SNR and spectral configuration. In Section~\ref{sec:discussion}, we discuss the implications of our findings in the context of previous studies and future mission design. Finally, Section~\ref{sec:conclusion} summarizes our conclusions and outlines directions for future work.

\section{Methods}\label{sec:methods}

We use \exorelr, a Bayesian atmospheric retrieval framework for spectroscopically resolved reflected-light observations of directly imaged exoplanets \citep{Damiano_2020,Damiano_2022,Damiano_2023,Tokadjian_2024}. A radiative-transfer forward model generates reflected-light spectra for a given set of planetary and atmospheric parameters. Model spectra are compared to a synthetic input spectrum, and nested sampling is used to infer posterior probability density distributions and Bayesian evidences.

Unless otherwise stated, all logarithms in this paper are base \num{10}.

\subsection{Upgrades to \mbox{\textsc{\textit{ExoReL$^\Re$}}}}\label{subsec:exorel_upgrades}

Relative to the version of \exorelr\ used in \citet{Damiano_2025}, some relevant model extensions have since been implemented. These upgrades are incorporated in this work and briefly described below.

\subsubsection{Wavelength-dependent surface albedo}\label{subsubsec:albedo}

The planetary surface is modeled as a Lambertian boundary characterized by a geometric surface albedo. Earlier \exorelr\ versions assumed a wavelength-independent (gray) surface albedo, but recent updates support a piecewise-constant albedo parameterization \citep[see also][]{Burr_2026}. In this work we adopt a two-step model with three albedo levels ($A_{\mathrm{surf},1}$, $A_{\mathrm{surf},2}$, $A_{\mathrm{surf},3}$) and two transition wavelengths. We retrieve the first transition wavelength $\lambda_{\mathrm{surf},1}$ and the positive step separation $\Delta_{\mathrm{surf}}$, with $\lambda_{\mathrm{surf},2}=\lambda_{\mathrm{surf},1}+\Delta_{\mathrm{surf}}$ by construction. This flexible but low-dimensional parameterization is intended to capture broad, wavelength-dependent reflectance behavior of heterogeneous surfaces without committing to a detailed physical surface model. For example, it can reproduce a generic rise in reflectance near the red/near-IR boundary that is qualitatively similar to the vegetation ``red edge'' (VRE) discussed in the context of Earthshine and exoplanet reflected-light spectra \citep[e.g.,][]{Arnold_2002,Woolf_2002,Seager_2005,MontanesRodriguez_2006,Kiang_2007}. However, we do not necessarily interpret the retrieved step as evidence for any specific surface composition; instead, it can also serve as a nuisance-parameter model to reduce biases in atmospheric inference arising from unknown surface spectral structure.

\subsubsection{$O_\mathit{2}$--$O_\mathit{2}$ CIA ($O_\mathit{4}$) opacities}\label{subsubsec:o2o2cia}

The \exorelr\ forward model now includes \OtwoOtwoCIA, which produces the \Ofour\ bands. \OtwoOtwoCIA\ arises from transient dipoles during \Otwo--\Otwo\ collisions and scales with the square of the \Otwo\ number density, making \Ofour\ particularly sensitive to atmospheric column density and composition.

\OtwoOtwoCIA\ is implemented using wavelength- and temperature-dependent coefficients from the HITRAN2020 CIA database \citep{Gordon_2022}, based on laboratory measurements and theoretical models \citep[e.g.,][]{Thalman_2013,Mondelain_2019,Karman_2018}. In the forward model, \OtwoOtwoCIA\ is treated as an additional continuum opacity alongside molecular line absorption, Rayleigh scattering, and cloud opacity.

\begin{figure*}[ht]
    \centering
    \includegraphics[width=\linewidth]{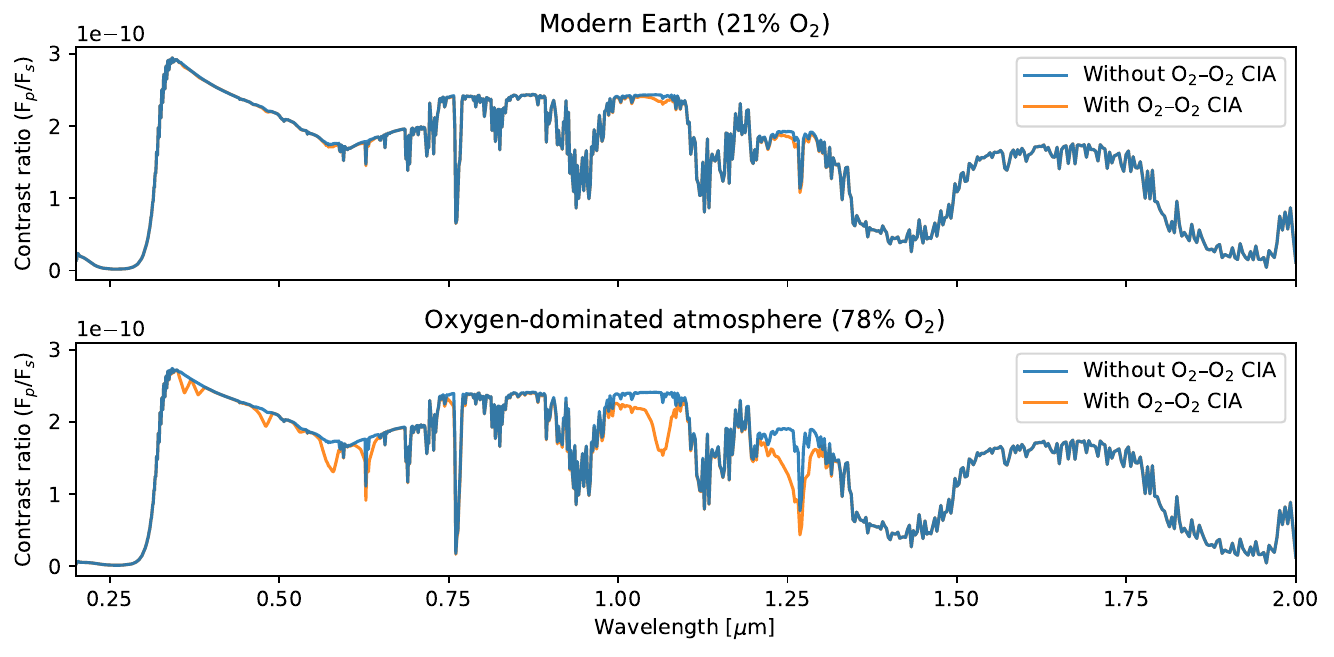}
    \caption{\OtwoOtwoCIA\ impact on reflected-light spectra. Top: modern Earth composition (\qty{21}{\percent} \Otwo). Bottom: otherwise identical \Otwo-dominated composition (\qty{78}{\percent} \Otwo), with \Ntwo\ adjusted to keep total surface pressure fixed.}
    \label{fig:cia_spectra}
\end{figure*}

Figure~\ref{fig:cia_spectra} illustrates the spectral impact of \OtwoOtwoCIA\ for two Earth-like compositions. \Ofour\ features are weak for a modern Earth \Otwo\ abundance and become pronounced for \Otwo-dominated compositions. The strongest \Ofour\ bands occur in the near-IR around \qtylist{1.06;1.27}{\micro\meter}. Additional bands appear across the near-UV/visible at \qtylist{0.34;0.36;0.38;0.45;0.48;0.53;0.58;0.63}{\micro\meter}, where they can overlap with the Rayleigh-scattering slope, and \OtwoOtwoCIA\ also makes a smaller contribution in the \Otwo\ A-band region near \qty{0.76}{\micro\meter}.

\subsection{Retrieval setup}\label{subsec:retrieval_setup}

The retrieval setup closely follows \citet{Damiano_2025} with the upgrades above. Atmospheric composition is parameterized by the (log) partial pressures of \HtwoO, \CHfour, \COtwo, \Otwo, \Othree, and \Ntwo, treated as independent parameters with log-uniform priors. For presentation, we convert retrieved partial pressures to volume mixing ratios (VMRs) when showing gas-abundance posteriors.

We include only water clouds, using the parameterization of \citet{Damiano_2020}. We retrieve the cloud-top pressure ($P_{\mathrm{top,H_2O}}$), cloud depth ($D_{\mathrm{cld,H_2O}}$), and condensation ratio ($\mathrm{CR}_{\mathrm{H_2O}}$).

Planetary mass and radius are treated as free parameters with uniform priors in all retrievals unless otherwise stated, enabling explicit tests of how \OtwoOtwoCIA\ reduces dependence on informative mass priors. The adopted bounds are intentionally broad nuisance priors rather than a population prior: they allow the retrieval to explore very weak external constraints on the size and gravity of a terrestrial-planet spectrum, while avoiding the clearly Jupiter-like regime. Because molecular hydrogen and helium are not included as atmospheric constituents in the forward model, posterior samples near the high-mass or high-radius edges of these priors should not be interpreted as physically self-consistent mini-Neptune or giant-planet models. Instead, these bounds are used to test whether the reflected-light data and \OtwoOtwoCIA\ can recover the correct Earth-like solution without imposing restrictive mass or radius information.

Table~\ref{tab:retrieval_setup} summarizes all retrieved parameters and priors. Sampling and Bayesian evidence calculation are performed with \texttt{MultiNest} \citep{Feroz_2009}, accessed through \texttt{pymultinest} \citep{Buchner_2014}. All retrievals use \num{1200} live points and an evidence tolerance of \num{0.5}.

\begin{deluxetable}{lcl}[ht]
\tablecaption{Retrieved parameters and prior distributions.\label{tab:retrieval_setup}}
\tablehead{\textbf{Parameter} & \textbf{Symbol} & \textbf{Prior}}
\startdata
Cloud-top pressure [\unit{\pascal}] & $P_{\mathrm{top,H_2O}}$ & $\mathcal{LU}(2,7)$ \\
Cloud depth [\unit{\pascal}] & $D_{\mathrm{cld,H_2O}}$ & $\mathcal{LU}(2,7)$ \\
Condensation ratio & $\mathrm{CR}_{\mathrm{H_2O}}$ & $\mathcal{LU}(-7,0)$ \\[3pt]
Partial pressure \HtwoO\ [\unit{\pascal}] & $P_{\mathrm{H_2O}}$ & $\mathcal{LU}(-7,7)$ \\
Partial pressure \CHfour\ [\unit{\pascal}] & $P_{\mathrm{CH_4}}$ & $\mathcal{LU}(-7,7)$ \\
Partial pressure \COtwo\ [\unit{\pascal}] & $P_{\mathrm{CO_2}}$ & $\mathcal{LU}(-7,7)$ \\
Partial pressure \Otwo\ [\unit{\pascal}] & $P_{\mathrm{O_2}}$ & $\mathcal{LU}(-7,7)$ \\
Partial pressure \Othree\ [\unit{\pascal}] & $P_{\mathrm{O_3}}$ & $\mathcal{LU}(-7,7)$ \\
Partial pressure \Ntwo\ [\unit{\pascal}] & $P_{\mathrm{N_2}}$ & $\mathcal{LU}(-7,7)$ \\[3pt]
Surface albedo (level 1) & $A_{\mathrm{surf},1}$ & $\mathcal{U}(0,1)$ \\
Surface albedo (level 2) & $A_{\mathrm{surf},2}$ & $\mathcal{U}(0,1)$ \\
Surface albedo (level 3) & $A_{\mathrm{surf},3}$ & $\mathcal{U}(0,1)$ \\
First albedo transition [\unit{\micro\meter}] & $\lambda_{\mathrm{surf},1}$ & $\mathcal{U}(0.4,0.8)$ \\
Step separation [\unit{\micro\meter}] & $\Delta_{\mathrm{surf}}$ & $\mathcal{U}(0.01,0.50)$ \\[3pt]
Planetary mass [\unit{\EarthMass}] & $M_p$ & $\mathcal{U}(0.01,19)$ \\
Planetary radius [\unit{\EarthRadius}] & $R_p$ & $\mathcal{U}(0.5,10.0)$ \\
\enddata
\tablecomments{$\mathcal{U}(a,b)$ denotes a uniform distribution between values $a$ and $b$, and $\mathcal{LU}(a,b)$ denotes a log-uniform distribution between values $a$ and $b$. The second albedo transition wavelength is defined as $\lambda_{\mathrm{surf},2} = \lambda_{\mathrm{surf},1} + \Delta_{\mathrm{surf}}$.}
\end{deluxetable}

\subsection{Atmospheric scenario and observational configurations}\label{subsec:scenarios}

All simulated observations adopt a modern Earth-like scenario: an \Ntwo-dominated atmosphere with \Otwo\ as a major constituent, trace \HtwoO, \Othree, \CHfour, and \COtwo, and a water cloud deck near \qty{\sim 0.6}{\bar} with a cloud fraction of \qty{25}{\percent}, unless otherwise stated. Planetary parameters are \qty{1}{\EarthMass}, \qty{1}{\EarthRadius}, and \qty{1}{\astronomicalunit} around a Sun-like star, with spectra computed at a phase angle of \qty{60}{\degree}. Input values are summarized in Table~\ref{tab:cia_comparison}.

To quantify the impact of \OtwoOtwoCIA\ and to assess SNR requirements for bulk-composition inference, we consider observational configurations motivated by recent HWO concept maturation and architecture studies \citep[e.g.,][]{Feinberg_2026}.

\textbf{Resolved-visible configuration.}
For the \OtwoOtwoCIA\ comparison and the SNR sweep, spectra are generated in the near-UV from \qtyrange{0.25}{0.43}{\micro\meter} at $\RNUV = \num{7}$, in the visible from \qtyrange{0.40}{1.00}{\micro\meter} at $\RVIS = \num{140}$, and in the near-IR from \qtyrange{1.0}{1.8}{\micro\meter} at $\RNIR = \num{70}$. For the \OtwoOtwoCIA\ comparison, we adopt $\mathrm{SNR} = \num{20}$. For the SNR sweep, we vary the SNR from \num{5} to \num{20}. As a sensitivity test, we repeat the retrievals with the near-IR resolution reduced to $\RNIR = \num{40}$, keeping the near-UV and visible settings fixed.

\textbf{Broadband-visible configuration.}
Another SNR study is conducted using a broadband-visible configuration in which the visible is summarized into three broad bins spanning \qtyrange{0.43}{0.52}{\micro\meter}, \qtyrange{0.50}{0.60}{\micro\meter}, and \qtyrange{0.58}{0.72}{\micro\meter}. In this case, spectra are generated in the near-UV from \qtyrange{0.25}{0.43}{\micro\meter} at $\RNUV = \num{7}$, from \qtyrange{0.70}{1.00}{\micro\meter} at $\RVIS = \num{140}$, and in the near-IR from \qtyrange{1.0}{1.7}{\micro\meter} at $\RNIR = \num{70}$. For the SNR sweep, we vary the SNR from \num{10} to \num{25}. As a sensitivity test, we repeat the retrievals with reduced near-IR wavelength coverage, truncating the near-IR range at \qty{1.4}{\micro\meter}.

\subsection{Noise model and retrieval strategy}\label{subsec:noise_strategy}

We adopt the noise model of \citet{Damiano_2025}, assuming photon-noise-limited observations with contributions from the planet signal and exozodiacal light; detector noise, local zodiacal light, and speckle residuals are neglected. The model is scaled such that the SNR at \qty{0.75}{\micro\meter} matches the specified value. We do not add random noise realizations to the synthetic spectra; instead, we use the noise model only to define wavelength-dependent uncertainties. This ``noise-free data with realistic error bars'' setup isolates the impact of model physics and parameter correlations from stochastic differences between individual noise draws, and thus approximates the noise-averaged behavior for a given SNR. For an individual observation, however, a particular noise realization can shift posterior modes, parameter medians, and credible intervals. As a check, we repeated the resolved-visible $\mathrm{SNR}=\num{20}$ retrieval for two Gaussian noise realizations drawn from the same wavelength-dependent uncertainties. Both noisy retrievals still recovered an \Ntwo-dominated atmosphere, but showed realization-dependent parameter biases; across these tests, most posterior shifts were at the \numrange{1}{2}$\sigma$ level, while the planet mass remained consistent with the input value within \num{3}$\sigma$.

We perform two sets of experiments. (i) To isolate the effect of \OtwoOtwoCIA, we run paired retrievals in which \OtwoOtwoCIA\ is either excluded or included consistently in both the synthetic spectrum and the retrieval forward model. (ii) To quantify observational requirements for robust bulk-composition inference, we perform SNR sweeps for the observing configurations described in Section~\ref{subsec:scenarios}. In addition to the Bayesian framework, we perform a simplified spectral separability test that compares forward-model spectra computed with and without \OtwoOtwoCIA\ across a grid of \Otwo\ abundances and SNR values, using a $\chi^2$-based metric over wavelength intervals containing prominent \OtwoOtwoCIA\ structure; this test is used to isolate the detectability of \OtwoOtwoCIA-induced spectral differences independent of retrieval degeneracies.

\section{Results}\label{sec:results}

We first assess the information content of \OtwoOtwoCIA\ for the modern Earth-like scenario. We then quantify the SNR required to resolve the \Ntwo-dominated versus \Otwo-dominated compositional degeneracy for the resolved-visible and broadband-visible configurations.

\subsection{Impact of $O_\mathit{2}$--$O_\mathit{2}$ CIA on reflected-light inference}\label{subsec:o4_results}

We present two complementary assessments of how forward-model completeness with respect to \OtwoOtwoCIA\ (i.e., the \Ofour\ bands) affects atmospheric inference for modern Earth-like scenarios. First, we perform a simplified spectral separability test that isolates the detectability of \OtwoOtwoCIA-induced spectral differences as a function of SNR and \Otwo\ abundance. Second, we compare full Bayesian retrievals performed without and with \OtwoOtwoCIA\ for an otherwise identical observing scenario.

\subsubsection{Spectral separability test}\label{subsubsec:o4_significance}

\begin{figure*}[ht]
    \centering
    \includegraphics[width=\linewidth]{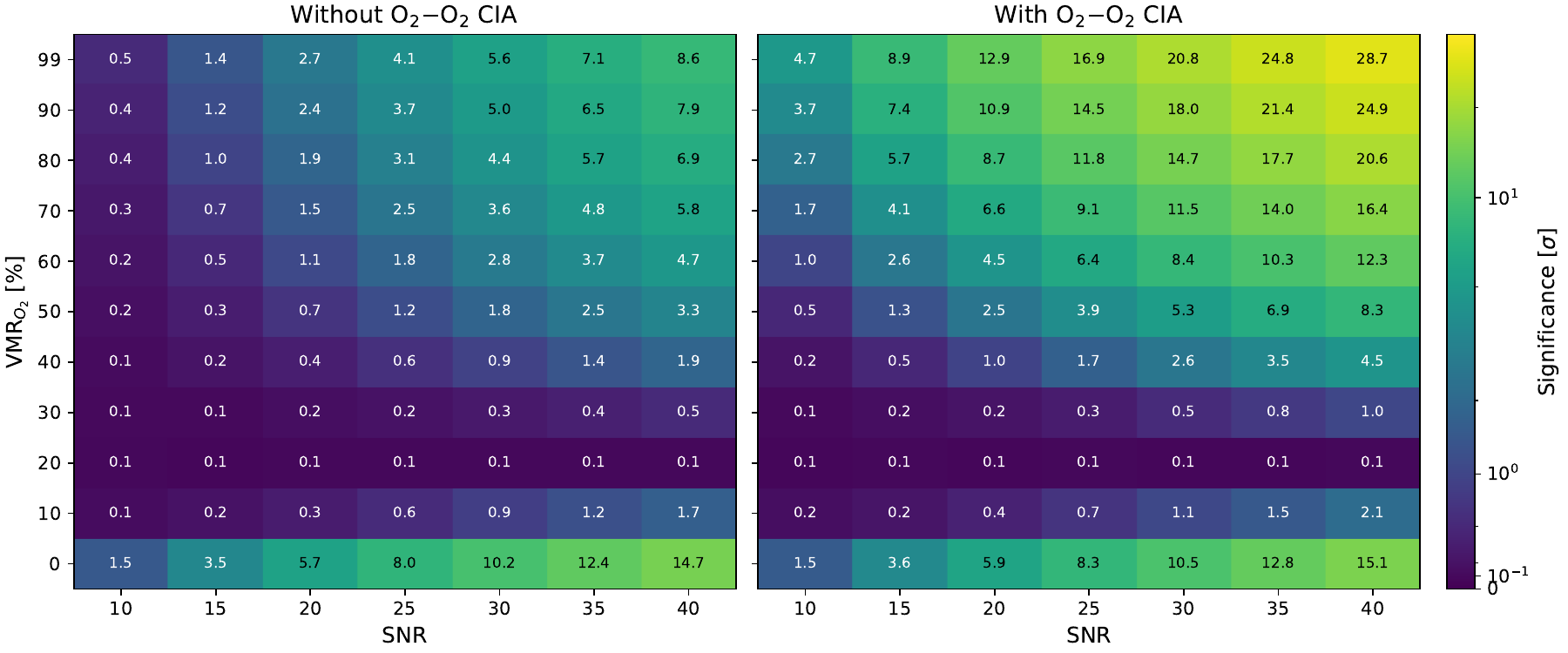}
    \caption{Mean significance for distinguishing spectra with different \Otwo\ VMRs from a \qty{20}{\percent} \Otwo\ reference spectrum. Left: \OtwoOtwoCIA\ excluded from model. Right: \OtwoOtwoCIA\ included in model.}
    \label{fig:cia_significance_heatmap}
\end{figure*}

As a simplified model-isolation test, we quantify how strongly \OtwoOtwoCIA\ alters reflected-light spectra as a function of \Otwo\ abundance and SNR. Using the \exorelr\ forward model, we generated spectra on a grid of $\mathrm{VMR}_{\Otwo}\in\{0,0.1,0.2,\dots,0.9,0.99\}$, adjusting \Ntwo\ to keep the total surface pressure fixed, and for reference $\mathrm{SNR}\in\{10,15,20,25,30,35,40\}$. For each grid point we computed a $\chi^2$ statistic relative to a reference spectrum $F_{\mathrm{ref}}(\lambda)$ with $\mathrm{VMR}_{\Otwo} = 0.2$, restricting the sum to near-IR intervals containing prominent \Otwo/\Ofour\ structure (\qtyrange{0.74}{0.78}{\micro\meter}, \qtyrange{0.97}{1.10}{\micro\meter}, and \qtyrange{1.20}{1.32}{\micro\meter}). Specifically, for each case we draw \num{1000} Gaussian-noise realizations $\tilde{F}_k(\lambda) \sim \mathcal{N}(F(\lambda),\sigma^2(\lambda))$ and compute
\begin{equation}
\chi^2_k \;=\; \sum_{\lambda\in\mathcal{I}}
\frac{\left[\tilde{F}_k(\lambda)-F_{\mathrm{ref}}(\lambda)\right]^2}{\sigma^2(\lambda)+\sigma_{\mathrm{ref}}^2(\lambda)} \, ,
\end{equation}
where $\mathcal{I}$ denotes the union of the intervals above, and $\sigma(\lambda)$ and $\sigma_{\mathrm{ref}}(\lambda)$ are the wavelength-dependent uncertainties from the adopted noise model at the specified SNR (added in quadrature because we compare two spectra with independent noise realizations). With $N$ data points in $\mathcal{I}$ and $\nu=N-1$ degrees of freedom, we compute the $p$-value using the cumulative distribution function,
\begin{equation}
p_k \;=\; 1 - \mathrm{CDF}_{\chi^2}\!\left(\chi^2_k;\nu\right),
\end{equation}
convert to an equivalent two-sided Gaussian significance,
\begin{equation}
Z_k \;=\; \mathrm{CDF}_{\mathcal{N}(0,1)}^{-1}\!\left(1-\frac{p_k}{2}\right),
\end{equation}
and report the mean $\langle Z\rangle$ over all noise realizations at each grid point. Figure~\ref{fig:cia_significance_heatmap} shows the resulting mean significances for the resolved-visible setup with $\RNIR = \num{70}$.

Including \OtwoOtwoCIA\ increases separability between \Ntwo-dominated, \Otwo-rich spectra and \Otwo-dominated spectra at high \Otwo\ abundances where \Ofour\ contributions are strongest. For example, when \OtwoOtwoCIA\ is omitted, a fixed-parameter comparison shows that the \qty{20}{\percent} \Otwo\ and \qty{99}{\percent} \Otwo\ cases are effectively indistinguishable at $\mathrm{SNR} = \num{20}$, consistent with reflected-light retrieval studies that can converge to \Otwo-dominated solutions for modern-Earth-like spectra under strong degeneracies \citep[e.g.,][]{Damiano_2025}; including \OtwoOtwoCIA\ introduces pronounced \Ofour\ structure that breaks this ambiguity. This test does not capture compensating parameter adjustments available in full retrievals, but it quantifies the intrinsic spectral leverage of \OtwoOtwoCIA\ at fixed model parameters.

We also assessed whether other CIA pairs involving the major gases could affect this conclusion. \Ntwo--\Ntwo\ CIA is included in the forward model but produces no appreciable spectral features over the \qtyrange{0.25}{1.8}{\micro\meter} range considered here. A forward-model test with \Ntwo--\Otwo\ CIA coefficients from HITRAN2020 \citep{Gordon_2022} shows only small contributions near the \Otwo\ A band at \qty{0.76}{\micro\meter} and the \Otwo/\Ofour\ band near \qty{1.27}{\micro\meter}. For the resolved-visible configuration at $\mathrm{SNR}=\num{20}$, including \Ntwo--\Otwo\ CIA changes the spectrum by at most \qty{5.55}{\percent} near \qty{1.264}{\micro\meter}, corresponding to \num{0.785}$\sigma$ in the most affected bin and a combined whole-spectrum mismatch of \num{0.902}$\sigma$ from the summed $\chi^2$ over all bins. Thus, \Ntwo--\Otwo\ CIA is below the assumed noise level in this setup and does not affect the retrieval conclusions, although it should be retained in future studies for completeness.

\subsubsection{Retrieval comparison}\label{subsubsec:o4_retrieval}

\begin{figure*}[ht]
    \centering
    \includegraphics[width=\linewidth]{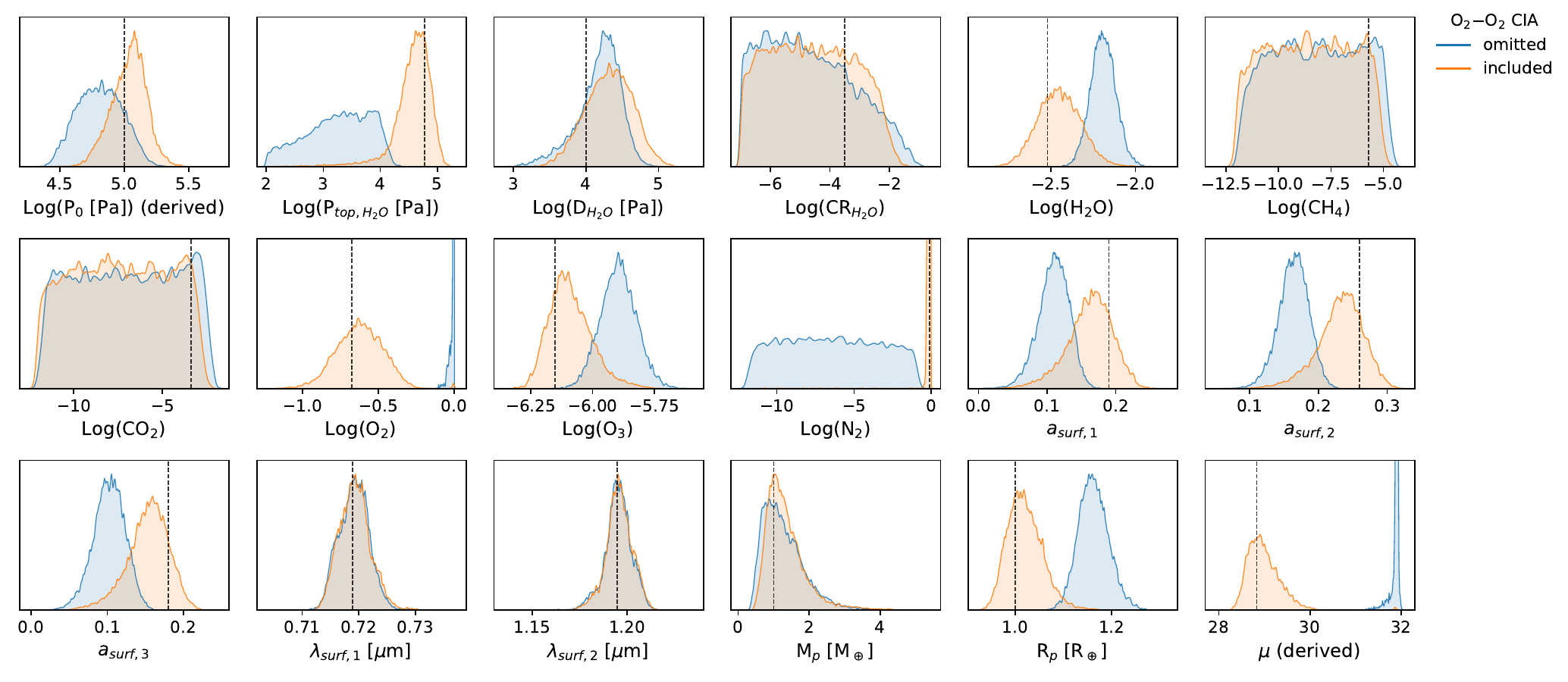}
    \caption{Posterior distributions for the modern Earth-like retrieval with \OtwoOtwoCIA\ omitted or included throughout (i.e., consistently omitted/included in both the synthetic input spectrum and the retrieval forward model). Dashed lines indicate input values.}
    \label{fig:posteriors_cia}
\end{figure*}

We next compare two otherwise identical retrievals that differ only in whether \OtwoOtwoCIA\ is excluded or included consistently in both the synthetic spectrum and the retrieval forward model (configuration: resolved-visible with $\RNIR = \num{70}$; $\mathrm{SNR} = \num{20}$). Posterior summaries are reported in Table~\ref{tab:cia_comparison}, and posteriors are compared in Figure~\ref{fig:posteriors_cia}. 

Without \OtwoOtwoCIA, the retrieval converges to an \Otwo-dominated solution with strongly suppressed \Ntwo. Including \OtwoOtwoCIA\ instead recovers an \Ntwo-dominated atmosphere consistent with the input model, which provides the correct bulk-composition baseline needed to interpret correlated parameters. In particular, once the dominant background gas (and thus the mean molecular mass) is correctly identified, the retrieval can attribute spectral features and continuum levels to physically consistent combinations of surface pressure, cloud-top pressure, radius, surface albedo, and correlated molecular abundances (e.g., \Othree) rather than compensating for an incorrect bulk composition. As a result, these coupled parameters shift toward their input values and become mutually consistent within the model. Notably, this change occurs despite the absence of an informative mass prior: in the \OtwoOtwoCIA-included case, the \Ofour\ bands provide enough internal spectral leverage to disfavor the spurious \Otwo-dominated mode that can otherwise persist when gravity is weakly constrained, in contrast to the strong mass-prior dependence reported when \OtwoOtwoCIA\ is not included \citep{Damiano_2025}.

To test whether mass prior knowledge can still further improve retrieval results, we additionally ran an \OtwoOtwoCIA-included retrieval with the planetary mass fixed to its true value. Fixing the mass does not significantly change the posteriors relative to the \OtwoOtwoCIA-included case with free mass, indicating that \Ofour\ alone provides sufficient information to favor the correct bulk composition for this scenario and data quality.

\begin{deluxetable}{lccc}[ht]
\tablecaption{Modern Earth-like retrieval comparison without and with \OtwoOtwoCIA\ (\Ofour).\label{tab:cia_comparison}}
\tablehead{
\textbf{Parameter} &
\textbf{Input} &
\textbf{Without $\mathrm{\mathbf{O_4}}$} &
\textbf{With $\mathrm{\mathbf{O_4}}$}
}
\startdata
Log($P_0$ [\unit{\pascal}]) (derived)
    & $5.00$
    & $4.81^{+0.18}_{-0.18}$
    & $5.05^{+0.12}_{-0.15}$ \\[3pt]
Log($P_{\mathrm{top,H_2O}}$ [\unit{\pascal}])
    & $4.78$
    & $3.31^{+0.55}_{-0.68}$
    & $4.62^{+0.22}_{-0.30}$ \\
Log($D_{\mathrm{cld,H_2O}}$ [\unit{\pascal}])
    & $4.00$
    & $4.23^{+0.23}_{-0.34}$
    & $4.33^{+0.32}_{-0.34}$ \\
Log(CR$_{\mathrm{H_2O}}$)
    & $-3.50$
    & $-4.83^{+1.84}_{-1.44}$
    & $-4.63^{+1.62}_{-1.54}$ \\[3pt]
Log(VMR$_{\mathrm{H_2O}}$)
    & $-2.52$
    & $-2.19^{+0.07}_{-0.07}$
    & $-2.45^{+0.13}_{-0.12}$ \\
Log(VMR$_{\mathrm{CH_4}}$)
    & $-5.70$
    & $-8.16^{+2.36}_{-2.34}$
    & $-8.53^{+2.31}_{-2.28}$ \\
Log(VMR$_{\mathrm{CO_2}}$)
    & $-3.40$
    & $-6.99^{+3.26}_{-3.27}$
    & $-7.36^{+3.13}_{-3.01}$ \\
Log(VMR$_{\mathrm{O_2}}$)
    & $-0.67$
    & $-0.003^{+0.001}_{-0.001}$
    & $-0.62^{+0.15}_{-0.15}$ \\
Log(VMR$_{\mathrm{O_3}}$)
    & $-6.15$
    & $-5.90^{+0.07}_{-0.07}$
    & $-6.10^{+0.09}_{-0.07}$ \\
Log(VMR$_{\mathrm{N_2}}$)
    & $-0.11$
    & $-6.51^{+3.68}_{-3.50}$
    & $-0.12^{+0.04}_{-0.06}$ \\[3pt]
$A_{\mathrm{surf},1}$
    & $0.19$
    & $0.11^{+0.02}_{-0.02}$
    & $0.16^{+0.03}_{-0.03}$ \\
$A_{\mathrm{surf},2}$
    & $0.26$
    & $0.16^{+0.02}_{-0.02}$
    & $0.24^{+0.03}_{-0.03}$ \\
$A_{\mathrm{surf},3}$
    & $0.18$
    & $0.10^{+0.02}_{-0.02}$
    & $0.16^{+0.02}_{-0.03}$ \\
$\lambda_{\mathrm{surf},1}$ [\unit{\micro\meter}]
    & $0.719$
    & $0.719^{+0.003}_{-0.003}$
    & $0.719^{+0.003}_{-0.003}$ \\
$\lambda_{\mathrm{surf},2}$ [\unit{\micro\meter}]
    & $1.195$
    & $1.195^{+0.007}_{-0.007}$
    & $1.196^{+0.007}_{-0.006}$ \\[3pt]
$M_p$ [\unit{\EarthMass}]
    & $1.00$
    & $1.17^{+0.65}_{-0.44}$
    & $1.23^{+0.58}_{-0.36}$ \\
$R_p$ [\unit{\EarthRadius}]
    & $1.00$
    & $1.16^{+0.03}_{-0.03}$
    & $1.02^{+0.04}_{-0.03}$ \\[3pt]
$\mu$ [\unit{\gram\per\mole}] (derived)
    & $28.83$
    & $31.91^{+0.02}_{-0.03}$
    & $28.93^{+0.39}_{-0.28}$ \\
\enddata
\tablecomments{Values are posterior medians with \qty{68}{\percent} credible intervals.}
\end{deluxetable}

\subsection{SNR requirement in the resolved-visible configuration}\label{subsec:snr_resolved}

We assess the SNR dependence of atmospheric inference for the resolved-visible configuration with $\RNIR = \num{70}$. Retrievals were performed for SNR values between \num{5} and \num{20}. Selected posteriors are shown in Figure~\ref{fig:posteriors_SNR_R70}, and the full posteriors for all parameters and SNR cases are provided in Appendix~\ref{app:snr_posteriors_r70}.

\begin{figure*}[ht]
    \centering
    \includegraphics[width=0.9\linewidth]{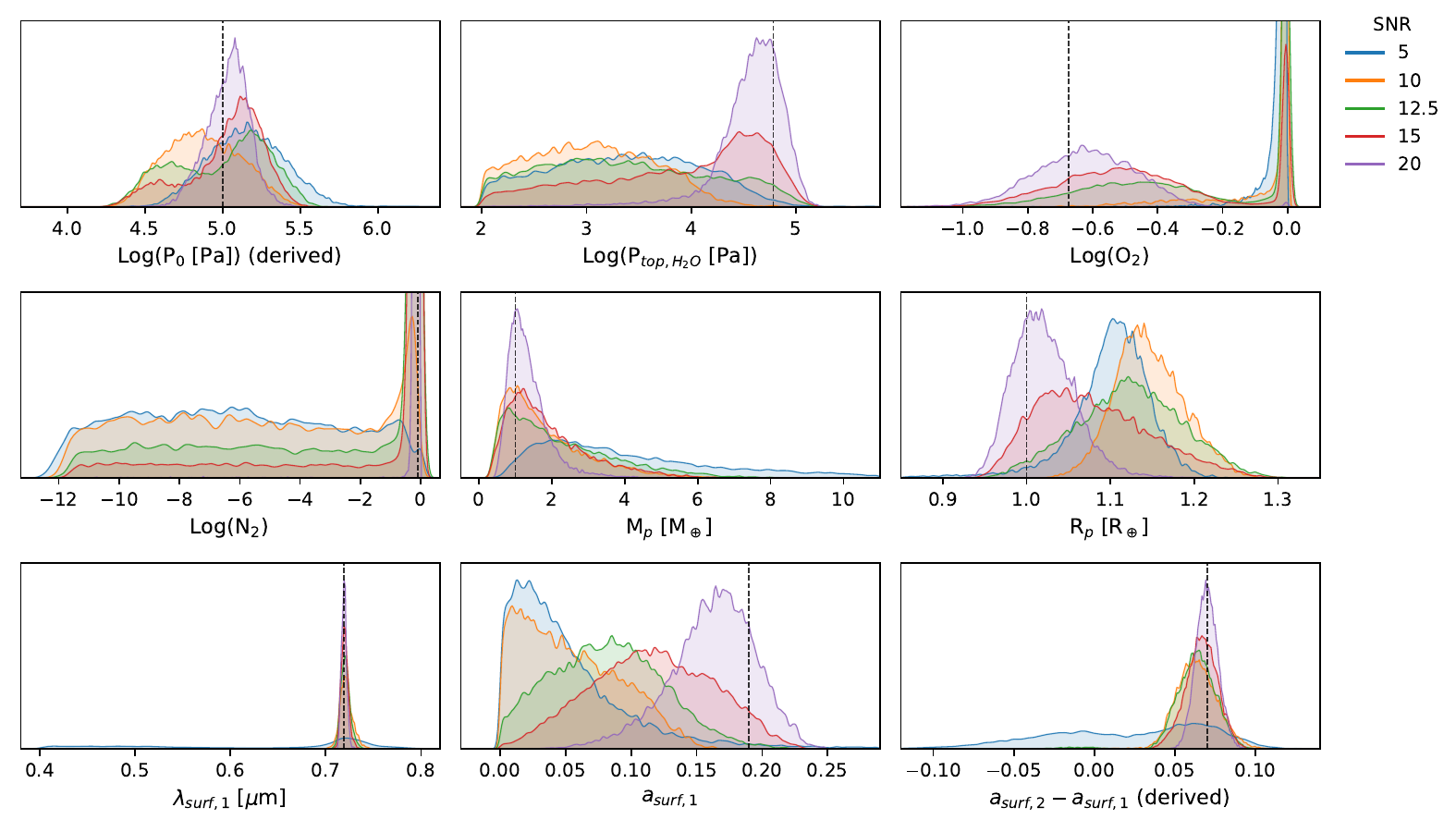}
    \caption{Selected posterior distributions for the SNR sweep in the resolved-visible configuration with $\RNIR = \num{70}$. Dashed lines indicate input values.}
    \label{fig:posteriors_SNR_R70}
\end{figure*}

At low SNR ($\lesssim \num{10}$), the posterior favors an \Otwo-dominated mode with strongly suppressed \Ntwo. With increasing SNR, the posterior shifts toward the correct \Ntwo-dominated solution and correlated parameters (surface pressure, albedo, and radius) tighten and move toward their input values.

\begin{figure}[ht]
    \centering
    \includegraphics[width=\figwidth]{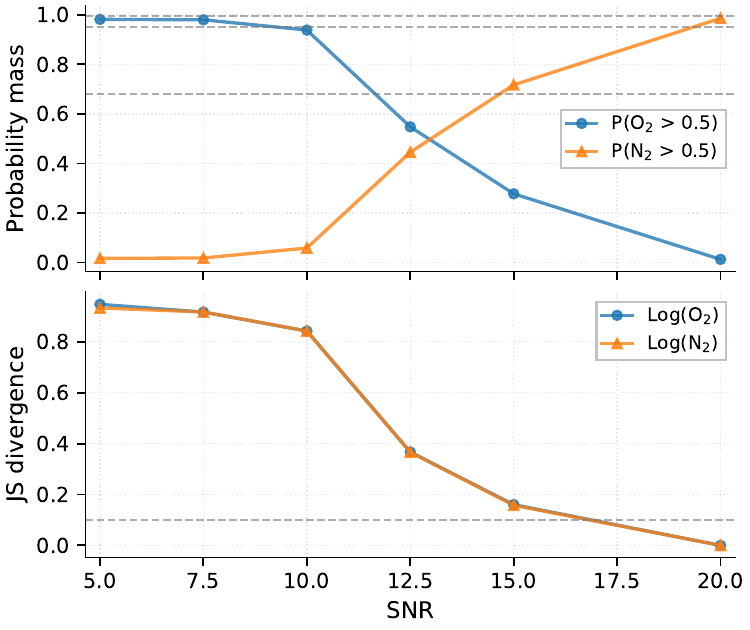}
    \caption{Bulk-composition versus SNR for the resolved-visible configuration ($\RNIR = \num{70}$). Top: posterior probability mass in \Ntwo-dominated and \Otwo-dominated solutions. The horizontal dashed lines mark heuristic posterior-mass thresholds at \qtylist{68;95;99.7}{\percent} for increasingly strong preference for the dominant gas. Bottom: Jensen--Shannon divergence relative to the $\mathrm{SNR} = \num{20}$ posteriors; the horizontal dashed line at $\mathrm{JS} = \num{0.1}$ marks an adopted heuristic threshold for strong distributional similarity.}
    \label{fig:dominant_gas_R70}
\end{figure}

Figure~\ref{fig:dominant_gas_R70} quantifies the evolution of the bulk-composition degeneracy as a function of SNR using two complementary metrics. The upper panel shows the posterior probability mass assigned to \Ntwo-dominated and \Otwo-dominated solutions as a function of SNR, computed by integrating the posterior over the regions where $\mathrm{VMR}_\Ntwo > \num{0.5}$ and $\mathrm{VMR}_\Otwo > \num{0.5}$, respectively. At intermediate SNR (\num{\sim 12.5}), both solutions coexist in the posterior, with the transition to a preferred \Ntwo-dominated solution occurring between $\mathrm{SNR} \approx \numlist{12.5;15}$. Using a more stringent criterion for robust dominant-gas identification (e.g., $\gtrsim \qty{95}{\percent}$ posterior mass in the \Ntwo-dominated region), the \Ntwo-dominated solution is only recovered at $\mathrm{SNR} \gtrsim \num{20}$ in this configuration.

The lower panel shows the Jensen--Shannon (JS) divergence \citep{Nielsen_2019} between the posterior distribution obtained at each SNR and the corresponding posterior at $\mathrm{SNR} = \num{20}$, which we adopt as a reference case. The JS divergence provides a symmetric, bounded measure of the similarity between two probability distributions, ranging from 0 (identical) to 1 (maximally different). Values below $\mathrm{JS}\approx \num{0.1}$ are often used heuristically to indicate strong distributional similarity, and the decreasing JS divergence with increasing SNR is consistent with the probability-mass analysis and supports the conclusion that $\mathrm{SNR} \gtrsim \num{20}$ is required for robust dominant-gas identification in this configuration.

The retrieved surface parameters exhibit two distinct behaviors. The location of the first albedo transition, $\lambda_{\mathrm{surf},1}$, is consistently recovered near its input value, and its posterior narrows steadily with increasing SNR. The albedo step amplitude $\Delta A_{\mathrm{surf}} \equiv A_{\mathrm{surf},2}-A_{\mathrm{surf},1}$ is clearly retrieved as positive, showing that a VRE-like reflectance rise near \qty{\sim 0.72}{\micro\meter} can be detected in this observing configuration. Only for $\mathrm{SNR} = 5$, there is a second mode with a negative step amplitude at lower wavelengths, but in that case the full posteriors reveal that the VRE is instead being described by $\lambda_{\mathrm{surf},2}$ and the corresponding albedo values. By contrast, the overall surface reflectance level (e.g., $A_{\mathrm{surf},1}$ as a proxy for the absolute normalization) is biased toward low values at low SNR, with the correct albedo level excluded for $\mathrm{SNR} \lesssim \num{12.5}$ in this configuration. The low surface reflectance is compensated by a larger radius, consistent with the radius--albedo scaling degeneracy, and the inferred absolute reflectance level becomes consistent with the input only as the dominant gas is correctly retrieved with increasing SNR.

To test sensitivity to near-IR resolution, we performed an otherwise identical sweep with $\RNIR = \num{40}$ (Appendix~\ref{app:snr_posteriors_r40}). Across the explored SNR range, we find no significant differences in the inferred posteriors or the bulk-composition transition.

Finally, we performed additional retrievals for a higher cloud fraction (\qty{50}{\percent} instead of \qty{25}{\percent}). In this case, the \Ntwo-dominated solution is recovered only at higher $\mathrm{SNR} \approx \num{25}$, and fixing the planetary mass accelerates recovery to $\mathrm{SNR} \approx \num{20}$. Posteriors are provided in Appendix~\ref{app:cloudfrac_posteriors}. In the case with fixed mass, the improved gravity constraint also reduces the radius--albedo degeneracy and yields more accurate posteriors for the absolute surface reflectance level.

\subsection{SNR requirement in the broadband-visible configuration}\label{subsec:snr_broadband}

We repeat the SNR sweep for the broadband-visible configuration (Section~\ref{subsec:scenarios}). Selected posteriors are shown in Figure~\ref{fig:posteriors_SNR_broadband}, and the full posteriors for all parameters are provided in Appendix~\ref{app:snr_posteriors_broadband}.

\begin{figure*}[ht]
    \centering
    \includegraphics[width=0.9\linewidth]{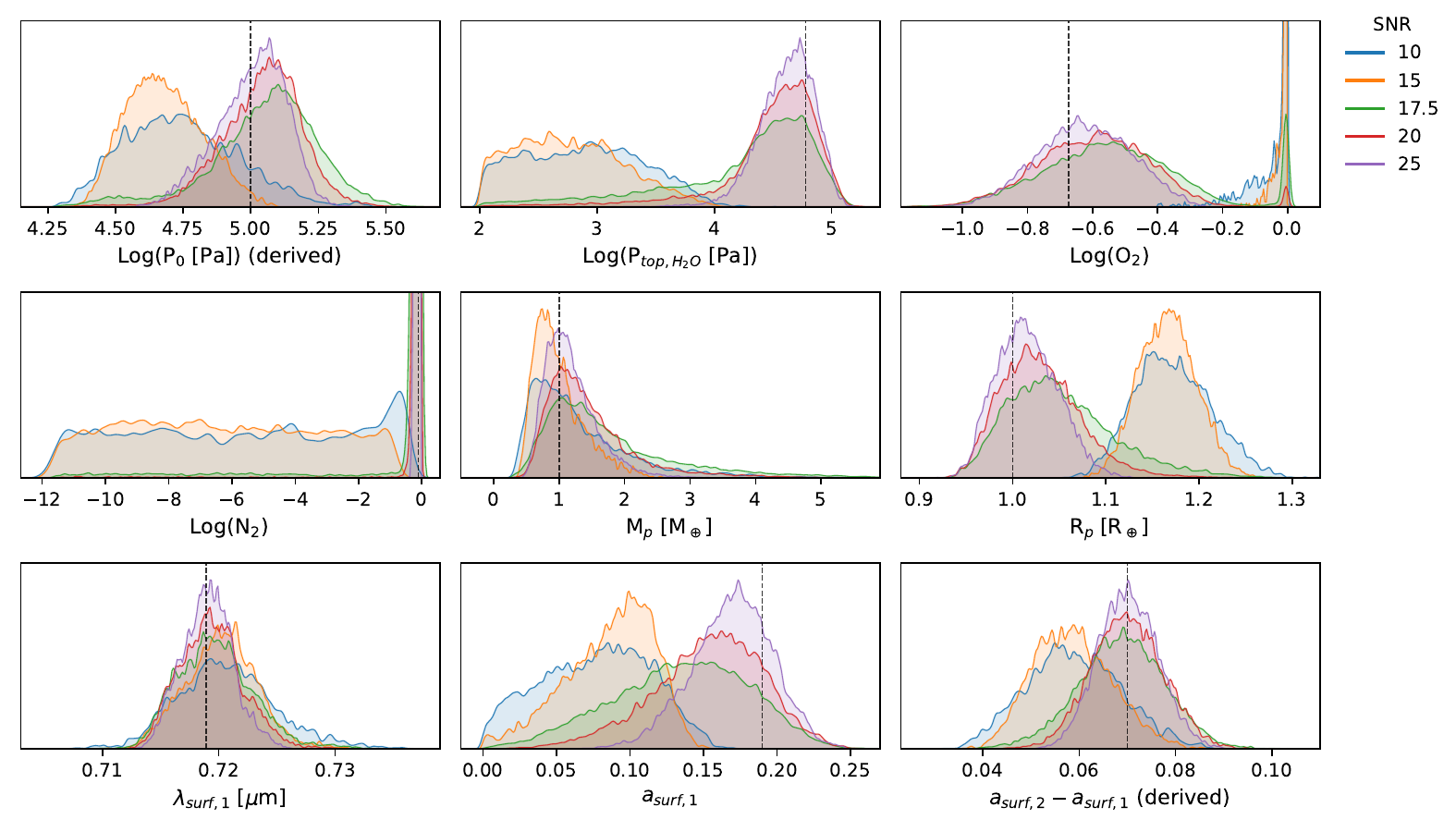}
    \caption{Selected posterior distributions for the SNR sweep in the broadband-visible configuration. Dashed lines indicate input values.}
    \label{fig:posteriors_SNR_broadband}
\end{figure*}

Similar to before, at low SNR ($\lesssim \num{15}$) the posterior favors an \Otwo-dominated solution with strongly suppressed \Ntwo\ and biases in correlated parameters. As SNR increases, the posterior shifts toward the correct \Ntwo-dominated solution.

\begin{figure}[ht]
    \centering
    \includegraphics[width=\figwidth]{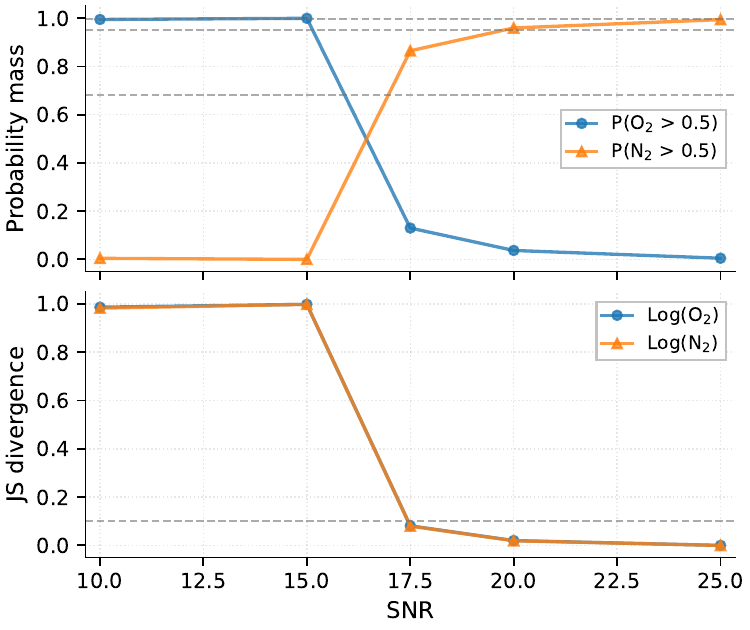}
    \caption{Bulk-composition versus SNR for the broadband-visible configuration. Top: posterior probability mass in \Ntwo-dominated and \Otwo-dominated solutions. The horizontal dashed lines mark heuristic posterior-mass thresholds at \qtylist{68;95;99.7}{\percent} for increasingly strong preference for the dominant gas. Bottom: Jensen--Shannon divergence relative to the $\mathrm{SNR} = \num{25}$ posteriors; the horizontal dashed line at $\mathrm{JS} = \num{0.1}$ marks an adopted heuristic threshold for strong distributional similarity.}
    \label{fig:dominant_gas_broadband}
\end{figure}

Figure~\ref{fig:dominant_gas_broadband} summarizes this evolution using posterior probability mass and JS divergence relative to the $\mathrm{SNR} = \num{25}$ posteriors. Compared to the resolved-visible case, the transition to a preferred \Ntwo-dominated solution occurs at higher SNR (between $\mathrm{SNR} \approx \numlist{15;17.5}$) when the visible is treated in broadband. However, a robust identification of the dominant gas (e.g., $\gtrsim \qty{95}{\percent}$ posterior mass in the \Ntwo-dominated region) is still achieved at $\mathrm{SNR} \gtrsim \num{20}$, which we adopt as the primary SNR requirement.

The JS divergence decreases with increasing SNR and approaches the high-SNR reference posterior. In this configuration the $\mathrm{JS} = \num{0.1}$ level is reached slightly earlier than the $\gtrsim \qty{95}{\percent}$ posterior-mass criterion (at $\mathrm{SNR} \gtrsim \num{17.5}$), but the two metrics are broadly consistent in indicating convergence toward the high-SNR solution.

The surface-albedo posteriors also show trends similar to the resolved-visible case. The first transition wavelength $\lambda_{\mathrm{surf},1}$ remains accurately constrained across the sweep and becomes more precise at higher SNR. The step amplitude $\Delta A_{\mathrm{surf}}=A_{\mathrm{surf},2}-A_{\mathrm{surf},1}$ is retrieved as positive for all SNR values considered here, showing that a VRE-like reflectance rise can be identified. As in the resolved-visible configuration, the absolute surface reflectance level is biased toward low values at low SNR and is entangled with the inferred radius; the correct $A_{\mathrm{surf},1}$ value is excluded for $\mathrm{SNR} \lesssim \num{15}$ in this configuration, and the posterior becomes consistent with the input albedo level only as the dominant gas is correctly retrieved at higher SNR.

\subsection{Broadband-visible configuration with reduced near-IR coverage}\label{subsubsec:snr_broadband_limited}

We additionally performed an SNR sweep for the same broadband-visible configuration but with the near-IR wavelength coverage truncated at \qty{1.4}{\micro\meter}. Selected posteriors are shown in Figure~\ref{fig:posteriors_SNR_broadband_limited}, and the full posteriors are provided in Appendix~\ref{app:snr_posteriors_broadband_limited}. Compared to the baseline broadband-visible case, reducing the near-IR coverage qualitatively changes the retrieval outcome: across the full explored SNR range from \numrange{20}{40}, the posterior remains in the spurious \Otwo-dominated mode and does not transition toward the correct \Ntwo-dominated solution.

\begin{figure*}[ht]
    \centering
    \includegraphics[width=0.9\linewidth]{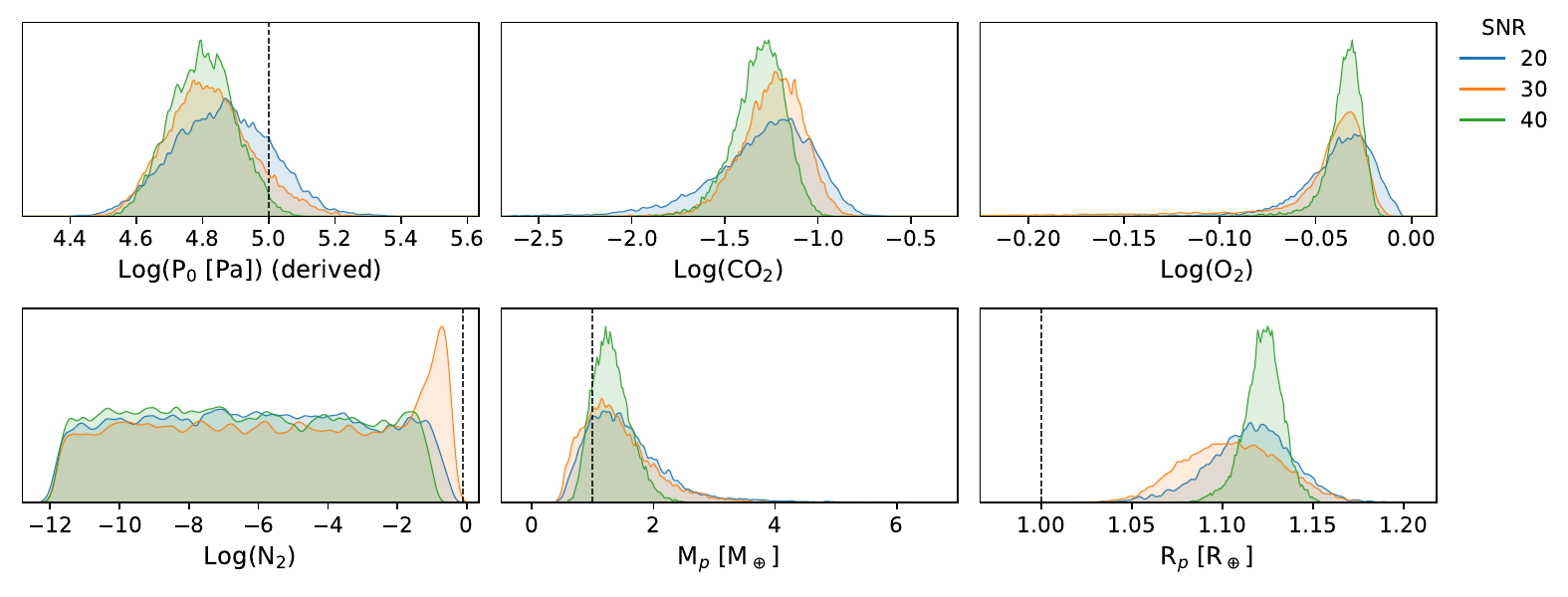}
    \caption{Selected posterior distributions for the SNR sweep in the broadband-visible configuration with reduced near-IR coverage (truncated at \qty{1.4}{\micro\meter}). Across the full explored SNR range, the retrieval remains in the spurious \Otwo-dominated mode. Dashed lines indicate input values.}
    \label{fig:posteriors_SNR_broadband_limited}
\end{figure*}

A notable difference from the broader-coverage broadband-visible runs is that the reduced-coverage retrievals also favor spuriously high \COtwo\ abundances, in addition to the biases already present in the incorrect \Otwo-dominated solution for \HtwoO, \Othree, and the cloud and surface parameters. Thus, removing the \qtyrange{1.4}{1.7}{\micro\meter} region does not merely weaken the constraints overall, but instead permits a specific compensating solution in which high \Otwo, elevated \COtwo, and correlated cloud and continuum adjustments jointly reproduce the observed spectrum.

\begin{figure*}[ht]
    \centering
    \includegraphics[width=\linewidth]{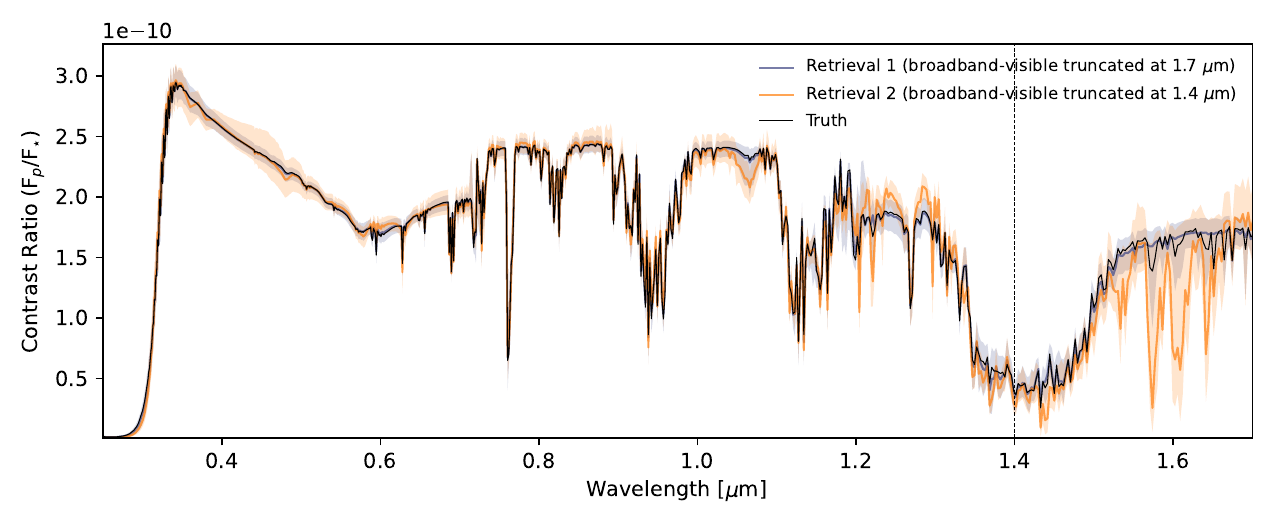}
    \caption{Comparison of retrieved spectra for the broadband-visible retrieval with near-IR coverage extending to \qty{1.7}{\micro\meter} and for the corresponding retrieval truncated at \qty{1.4}{\micro\meter}, both at $\mathrm{SNR} = \num{20}$. Solid lines show the median retrieved spectra, and the shaded regions indicate the \qtyrange{0.3}{99.7}{\percent} quantile interval from the posterior samples. The true input spectrum is overplotted at equally high spectral resolution to make the detailed structure more visible. Although the truncated-coverage retrieval is fit only below \qty{1.4}{\micro\meter}, its posterior-predictive spectrum is shown out to \qty{1.7}{\micro\meter}; the strong disagreement with the truth at longer wavelengths reveals the cost of removing that spectral coverage.}
    \label{fig:broadband_coverage_spectra}
\end{figure*}

To illustrate this effect, Figure~\ref{fig:broadband_coverage_spectra} compares the median retrieved spectra for the broadband-visible retrieval with near-IR coverage extending to \qty{1.7}{\micro\meter} and for the corresponding retrieval truncated at \qty{1.4}{\micro\meter}, both at $\mathrm{SNR} = \num{20}$, together with the true spectrum shown at high spectral resolution. Over the observed wavelength range below \qty{1.4}{\micro\meter}, both retrievals reproduce the truth reasonably well, with visible differences mainly around the \Ofour-sensitive regions. When the spectra are viewed over the extended \qtyrange{1.4}{1.7}{\micro\meter} interval, however, the disagreement becomes much stronger: the reduced-coverage retrieval predicts substantially stronger absorption, most notably in the longer-wavelength \COtwo\ features. The missing long-wavelength near-IR data are therefore crucial for rejecting the incorrect \Otwo-dominated, \COtwo-rich solution.

\begin{figure*}[ht]
    \centering
    \includegraphics[width=0.8\linewidth]{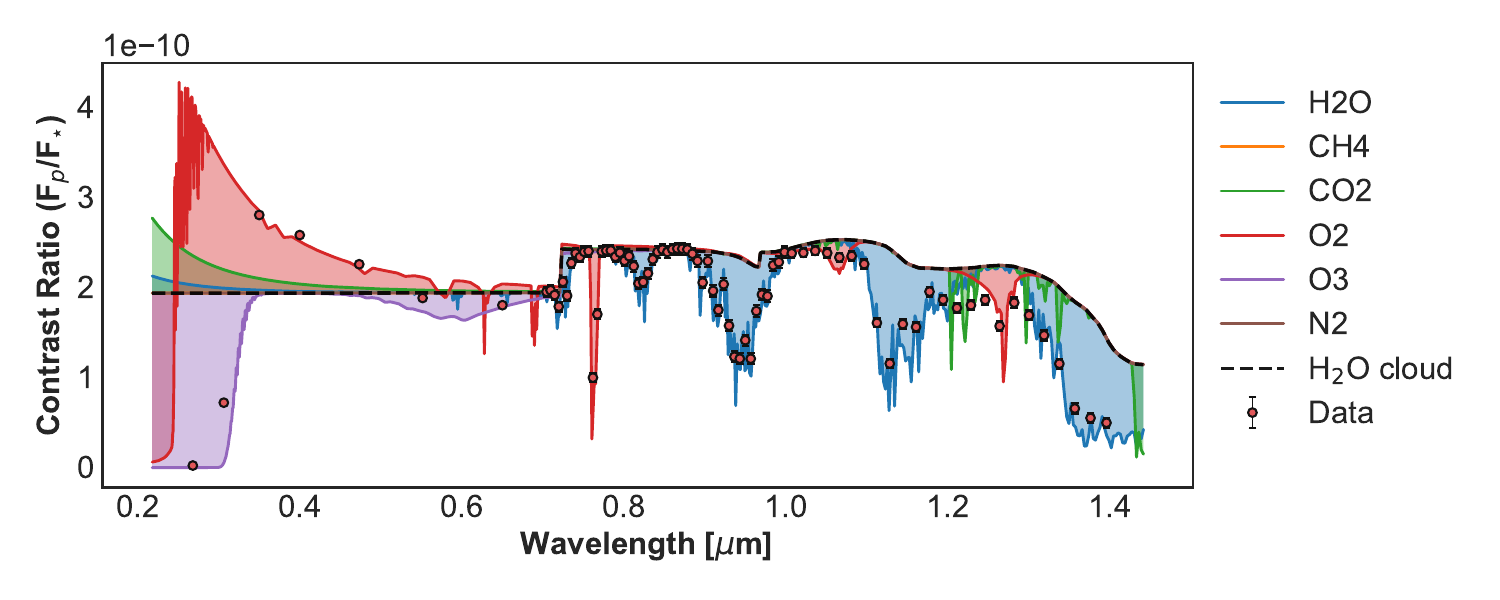}
    \caption{Molecular contributions to the maximum-a-posteriori spectrum for the broadband-visible retrieval with near-IR coverage truncated at \qty{1.4}{\micro\meter} at $\mathrm{SNR} = \num{40}$. The dashed black curve shows the baseline spectrum obtained with molecular opacities removed, retaining only cloud scattering and surface reflectance. Colored curves show the spectrum after restoring the opacity of one molecule at a time on top of that baseline, thereby isolating how each species modifies the reflected-light spectrum. The data points are the synthetic observations.}
    \label{fig:broadband_mol_contrib_limited}
\end{figure*}

The molecular-contribution view in Figure~\ref{fig:broadband_mol_contrib_limited} helps explain how the reduced-coverage retrieval reproduces the data despite converging to an incorrect solution. The figure shows the maximum-a-posteriori spectrum for the $\mathrm{SNR} = \num{40}$ reduced-coverage retrieval, decomposed relative to a baseline spectrum that includes only cloud scattering and surface reflectance. In the \qtyrange{1.2}{1.3}{\micro\meter} region, the baseline continuum is retrieved too high, and the retrieval compensates by increasing \COtwo\, which has weaker absorption features in the same range. The figure also shows that, in addition to the dominant \Otwo, \COtwo\ contributes noticeably to the near-UV continuum through Rayleigh scattering.

The surface-albedo parameters still indicate a detectable VRE-like rise across the explored SNR range. At the same time, the second albedo step (i.e., $\lambda_{\mathrm{surf},2}$) is not correctly recovered in these reduced-coverage runs. This indicates that the VRE location near \qty{\sim 0.72}{\micro\meter} remains well identified, while the placement of the other albedo transition is less reliably constrained and can be retrieved at other wavelengths, a behavior also seen for some solutions in the other configurations.

\section{Discussion}\label{sec:discussion}

\subsection{Context relative to previous reflected-light retrieval work}\label{subsec:discussion_prev}

Our results reinforce a central conclusion of reflected-light retrieval studies for temperate terrestrial planets: bulk composition and atmospheric column remain weakly constrained unless the observations provide sufficient information content (through SNR, spectral resolution, and wavelength coverage) and/or informative external constraints on system geometry and bulk properties are available \citep[e.g.,][]{Carrion_Gonzalez_2020,Robinson_2023,Salvador_2025}. This reflects the fact that reflected-light spectra are shaped by coupled effects of composition (molecular bands and Rayleigh scattering), column abundance (surface pressure and gravity), and continuum-controlling parameters (clouds and surface reflectance). Prior work has emphasized, in particular, that uncertain radius and illumination geometry can introduce strong degeneracies with albedo and cloud properties, motivating multi-epoch or multi-phase strategies and the use of external constraints whenever possible \citep[e.g.,][]{Nayak_2017,Carrion_Gonzalez_2021,Salvador_2024}.

Within this broader context, the emergence of competing \Ntwo-dominated and \Otwo-dominated solutions at modest SNR in our \exorelr\ retrievals is consistent with earlier terrestrial reflected-light retrieval studies that report multimodality and composition--column--continuum degeneracies, especially when radius and gravity are not tightly constrained \citep[e.g.,][]{Feng_2018,Damiano_2022,Damiano_2025}. Practically, detecting \Otwo\ features does not by itself determine whether an atmosphere is \Ntwo-dominated and oxygen-rich, or \Otwo-dominated; robust inference depends on whether the data constrain the atmospheric column and the continuum-shaping parameters (clouds and surface reflectance) well enough to prevent compensating solutions.

Relative to this previous work, the key difference explored here is the explicit inclusion of \OtwoOtwoCIA\ in the reflected-light retrieval forward model and its interplay with observing configuration. We find that including \OtwoOtwoCIA\ is not a minor correction: it can qualitatively reshape the posterior by disfavoring high-\Otwo\ column solutions that can otherwise reproduce the monomer \Otwo\ bands via correlated adjustments in surface pressure, radius, clouds, and surface albedo (Section~\ref{subsec:o4_results}, Figure~\ref{fig:posteriors_cia}). This complements external constraints: for example, orbit/radius information restricts the allowable scaling between albedo and radius, and mass (or gravity) information restricts the mapping between pressure and column, while \OtwoOtwoCIA\ adds an internal, spectrally encoded constraint on the \Otwo\ column density. The extent to which this internal spectral leverage can be translated into a robust dominant-gas identification depends on the total information content of the observations, including spectral resolution, wavelength coverage, and SNR.

\subsection{$O_\mathit{2}$--$O_\mathit{2}$ CIA as a discriminator of bulk composition}\label{subsec:discussion_o4}

The physical reason \OtwoOtwoCIA\ improves discrimination between \Ntwo-dominated, oxygen-rich atmospheres and \Otwo-dominated atmospheres is its nonlinear dependence on \Otwo\ number density: the strength of the associated \Ofour\ bands scales approximately with the line-of-sight integral of $n_\Otwo^2$ \citep[e.g.,][]{Greenblatt_1990,Karman_2018}. In a retrieval, this provides leverage distinct from the approximately linear dependence of monomer O$_2$ bands on \Otwo\ abundance. If the fit attempts to explain spectral structure by increasing the \Otwo\ partial pressure (or the total column), the \Ofour\ bands strengthen rapidly and can overpredict absorption in \Ofour-sensitive regions, thereby suppressing otherwise viable \Otwo-dominated modes.

This mechanism is most evident in the moderate-SNR regime where the posterior is multimodal. In that regime, one family of solutions corresponds to an \Ntwo-dominated atmosphere with Earth-like \Otwo\ that matches the continuum, while a second family corresponds to an \Otwo-dominated atmosphere that compensates through correlated shifts in surface pressure, radius, cloud properties, and surface albedo. Including \OtwoOtwoCIA\ enforces consistency between monomer O$_2$ bands and the \Ofour\ bands, shifting posterior mass toward the physically correct \Ntwo-dominated solution when the data quality and wavelength sampling are sufficient to detect \Ofour\ structure. The wavelength distribution of \Ofour\ information is also important for how this leverage survives instrument binning: the strong near-IR \Ofour\ bands (near 1.06 and 1.27~\unit{\micro\meter}) are both deeper and broader at high \Otwo\ columns, and therefore remain detectable at modest resolving power, whereas the visible \Ofour\ contribution near \qty{\sim 0.63}{\micro\meter} is comparatively narrow and can be more easily diluted when averaged into wide bins. By contrast, the monomer \Otwo\ A-band is already very deep at modern-Earth \Otwo\ and is comparatively narrow, so its incremental utility for discriminating bulk composition decreases when the visible is coarsely binned.

The interaction with planet-mass information is complementary. Mass constraints reduce flexibility in trading off surface gravity against pressure and composition \citep{Damiano_2025}. \OtwoOtwoCIA\ constrains the \Otwo\ column spectroscopically, limiting how much of the atmospheric column can be assigned to \Otwo\ versus the background gas. When both are available, inference becomes more robust to model flexibility in clouds and surface reflectance; when mass information is weak, \OtwoOtwoCIA\ can partially compensate, but only if the \Ofour-sensitive regions are observed at adequate SNR. In practical terms, this added spectral leverage could relax the requirement that the planet mass be known to within \qty{\sim 10}{\percent} for modern Earth--like atmospheres in which \Otwo\ is a major constituent. However, these effects cannot be exploited when the atmosphere contains only a small amount of \Otwo\ and is instead dominated by other gases. In that regime, the \OtwoOtwoCIA\ signal is intrinsically weak, because too little \Otwo\ is present for \Ofour\ absorption to provide a meaningful constraint on the \Otwo\ column or on how the total atmospheric column is partitioned among atmospheric constituents. As a result, \OtwoOtwoCIA\ offers limited leverage for compensating for weak external mass constraints, even if the \Ofour-sensitive regions are observed at high SNR. Archean Earth--like atmospheres provide a useful example of this regime \citep[e.g.,][]{Damiano_2025}: because \Otwo\ is negligible, \OtwoOtwoCIA\ contributes little to no additional information, and external mass constraints remain necessary. More generally, when \Otwo\ is not available as a major spectroscopic handle, the dominant bulk-composition degeneracies depend on which other gases make up the atmosphere and how their spectral signatures overlap. For Archean-like cases, for example, the principal trade-off may occur between \Ntwo\ and \COtwo; once the mass uncertainty exceeds \qty{\sim 10}{\percent}, a broad range of such compositions can remain consistent with the data.

\subsection{Implications for observing strategy and spectral configuration}\label{subsec:discussion_obs}

For mission-relevant observing requirements, the relevant metric is the SNR at which it is recovered with high confidence. Adopting a robustness criterion of \qty{\geq 95}{\percent} posterior mass in the \Ntwo-dominated region, both the resolved-visible and broadband-visible configurations require $\mathrm{SNR}\approx \num{20}$ in our setup to robustly favor the correct \Ntwo-dominated atmosphere. Thus, although the two configurations exhibit different onset behavior (the posterior transitions from \Otwo-dominated to \Ntwo-dominated at lower SNR in the resolved-visible case than in the broadband-visible case), the SNR requirement for a robust dominant-gas identification is essentially the same for both.

The differing onset behavior is consistent with how the visible continuum is sampled. In our setup, the primary difference between the two configurations is the treatment of the blue visible, where the Rayleigh-scattering slope carries key continuum information that helps constrain atmospheric column and background gas. When these wavelengths are spectrally resolved, the retrieval retains leverage on the wavelength-dependent Rayleigh continuum (and its coupling to cloud-top pressure and surface reflectance), which causes the \Ntwo-dominated mode to emerge earlier in SNR. When the same region is compressed into broad photometric bins, much of this Rayleigh leverage is lost, allowing compensating solutions to persist to higher SNR. However, eliminating the competing \Otwo-dominated mode at the \qty{\sim 95}{\percent} level still requires sufficiently small uncertainties across the continuum and \Ofour-sensitive regions, which in our experiments occurs only once $\mathrm{SNR}\approx \num{20}$ is reached.

This threshold should be interpreted within the limitations of the adopted noise model. In particular, our tests assume independent Gaussian errors and do not include spectrally correlated noise, such as residual speckle correlations. Recent HWO-focused retrieval work has shown that such correlations can broaden or bias retrieved molecular-abundance and planetary-parameter posteriors \citep{Wolff_2026}, so incorporating correlated noise would be an important extension of the present observing-strategy analysis.

Our reduced near-IR resolution test indicates that lowering the near-IR resolving power from $\RNIR = \num{70}$ to $\RNIR = \num{40}$ does not significantly change retrieval behavior across the explored SNR range (Appendix~\ref{app:snr_posteriors_r40}). This suggests that modest near-IR resolution changes are not the dominant driver of the robust-composition threshold for this scenario, and that the principal difference between the resolved-visible and broadband-visible configurations is the degree to which the near-UV/blue continuum information (including the Rayleigh slope) is preserved.

By contrast, reducing near-IR wavelength coverage can have a much stronger impact. Truncating the near-IR range from \qty{1.7}{\micro\meter} to \qty{1.4}{\micro\meter} in the broadband-visible configuration leaves the retrieval trapped in a spurious \Otwo-dominated mode across the full explored SNR range (up to $\mathrm{SNR}=\num{40}$). In these reduced-coverage runs, the posterior also favors spuriously high \COtwo\ abundances, together with broader correlated adjustments in cloud and continuum-setting parameters. This suggests that the omitted \qtyrange{1.4}{1.7}{\micro\meter} interval provides critical spectral context that is needed to reject the compensating \Otwo-dominated, \COtwo-rich solution. In other words, retaining coverage beyond \qty{1.4}{\micro\meter} is important not just for lowering the SNR requirement, but for enabling the correct bulk-composition inference at all within this observing configuration. Quantifying which specific spectral intervals carry the most leverage in this regime, and how this depends on cloud and surface assumptions, would be a natural extension of this work.

We also performed supplemental reduced-coverage retrievals with stronger radius information (Appendix~\ref{app:radius_prior_posteriors}). With a Gaussian radius prior centered on the input value with \qty{10}{\percent} standard deviation, the posterior begins to recover the correct \Ntwo-dominated solution but remains degenerate. When the radius is fixed to its input value, the retrieval strongly favors the \Ntwo-dominated mode. In both radius-constrained cases, the \COtwo\ posterior no longer concentrates at the spuriously high abundance found with an unconstrained radius; instead, high \COtwo\ abundances are excluded while lower abundances remain weakly constrained.

\subsection{Sensitivity to clouds and surface reflectance assumptions}\label{subsec:discussion_clouds_albedo}

Clouds and surfaces affect reflected-light spectra primarily by reshaping the continuum and modifying the effective photon path length through absorbing layers. Increasing cloud fraction generally reduces sensitivity to deeper atmospheric layers and weakens spectral contrast by raising and flattening the reflectance baseline. Consistent with this intuition, our higher cloud-fraction tests exhibit stronger degeneracies and require higher SNR to robustly recover the \Ntwo-dominated solution. Fixing the planetary mass partially mitigates this effect, indicating that external gravity constraints remain valuable in more challenging cloud regimes (Appendix~\ref{app:cloudfrac_posteriors}; \citealt{Damiano_2025}).

Surface reflectance introduces additional degeneracies by modulating the broadband continuum against which molecular features are measured, and by coupling directly to the absolute flux level through the familiar $(R_p/a)^2 A_g(\lambda)$ scaling. As a result, reflected-light retrievals commonly exhibit a radius--albedo degeneracy that can persist even at relatively high SNR when the surface spectrum is flexible \citep[e.g.,][]{Nayak_2017,Carrion_Gonzalez_2020,Carrion_Gonzalez_2021}. Compared to \citet{Damiano_2025}, who adopted a constant surface albedo of 0.05, we use a wavelength-dependent (step-function) albedo with values of order \num{\sim 0.2}, corresponding to a different disk-integrated Earth realization and enabling broad spectral structure to be absorbed by the surface model rather than being forced into atmospheric parameters. This choice is also motivated by recent work emphasizing that heterogeneous surfaces and clouds can imprint wavelength-dependent signatures in reflected light and can impact the retrievability of surface features such as a red-edge-like reflectance rise \citep{Burr_2026}.

In our SNR sweeps, the first step location parameter ($\lambda_{\mathrm{surf},1}$) is consistently well constrained and tightens with increasing SNR, while the retrieved step direction decisively finds $A_{\mathrm{surf},2} > A_{\mathrm{surf},1}$, indicating that the retrieval recognizes a red/near-IR reflectance rise within this generic parameterization. At the same time, the absolute surface reflectance level remains more difficult to recover and is strongly coupled to $R_p$, with low-SNR cases favoring lower albedo and larger radius solutions. This reinforces that even when the overall surface reflectance level is biased by the persistent radius--albedo scaling degeneracy, the spectral shape of the surface can remain identifiable.

\section{Conclusion}\label{sec:conclusion}

We used \exorelr\ to quantify how forward-model completeness and observing configuration affect reflected-light constraints on the bulk composition of a modern Earth analog, with the specific goal of distinguishing \Ntwo-dominated (\Otwo-rich) from \Otwo-dominated atmospheres. Two model upgrades are central: inclusion of \OtwoOtwoCIA\ (producing \Ofour\ bands), and a wavelength-dependent step-function surface albedo.

Including \OtwoOtwoCIA\ qualitatively changes bulk-composition inference. Otherwise identical retrievals omitting \OtwoOtwoCIA\ converge to an incorrect \Otwo-dominated solution, while retrievals including \OtwoOtwoCIA\ recover the correct \Ntwo-dominated atmosphere. In the \OtwoOtwoCIA-included case, fixing the planetary mass produces no significant change in the posteriors for the nominal scenario, indicating reduced reliance on tight mass priors for modern Earth--like atmospheres in which \Otwo\ is a major constituent.

SNR requirements depend on spectral information content and on the robustness criterion adopted. Using a requirement of \qty{\geq 95}{\percent} posterior mass in the \Ntwo-dominated region, both the resolved-visible and broadband-visible configurations with near-IR coverage to \qty{1.7}{\micro\meter} require $\mathrm{SNR}\approx \num{20}$ to confidently identify the correct dominant gas in our setup. Higher cloud fraction delays robust recovery to higher SNR, and informative mass priors remain beneficial in that regime.

Near-IR wavelength coverage can materially affect whether the correct bulk composition is recovered at all. In an otherwise identical broadband-visible configuration with the near-IR truncated at \qty{1.4}{\micro\meter}, the retrieval remains trapped in a spurious \Otwo-dominated mode across the full explored SNR range (up to $\mathrm{SNR}=\num{40}$) and also favors spuriously high \COtwo\ abundances together with broader correlated adjustments in cloud and continuum-setting parameters. This sensitivity highlights the importance of retaining near-IR bands that constrain correlated absorbers such as \COtwo\ and help suppress compensating \Otwo-dominated, \COtwo-rich solutions in bulk-composition retrievals.

Overall, these results strengthen the Habitable Worlds Observatory science case by clarifying how retrieval-model completeness and instrument choices (SNR, spectral sampling, and wavelength coverage) govern the robustness of oxygen interpretation for potentially habitable, Earth-like planets.

\begin{acknowledgments}
We thank Armen Tokadjian for contributions to the development of \exorelr, in particular for implementing \OtwoOtwoCIA\ in the forward model.
The High Performance Computing resources used in this investigation were provided by funding from the JPL Information and Technology Solutions Directorate.
This research was carried out at the Jet Propulsion Laboratory, California Institute of Technology, under a contract with the National Aeronautics and Space Administration (80NM0018D0004).
\end{acknowledgments}

\software{Most retrievals in this work used \exorelr\ version~2.7.0. The supplemental noisy-data checks in Section~\ref{subsec:noise_strategy} and reduced-coverage radius-prior retrievals in Section~\ref{subsec:discussion_obs} used \exorelr\ version~3.1.2. \exorelr\ is an open-source project and is available on \href{https://github.com/MDamiano/ExoReL}{GitHub}.}

\appendix
\section{Additional posterior distributions}\label{app:posteriors}

\subsection{SNR sweep posteriors: resolved-visible ($\mathit{R_{NIR} = 70}$)}\label{app:snr_posteriors_r70}

\begin{figure*}[ht]
    \centering
    \includegraphics[width=\linewidth]{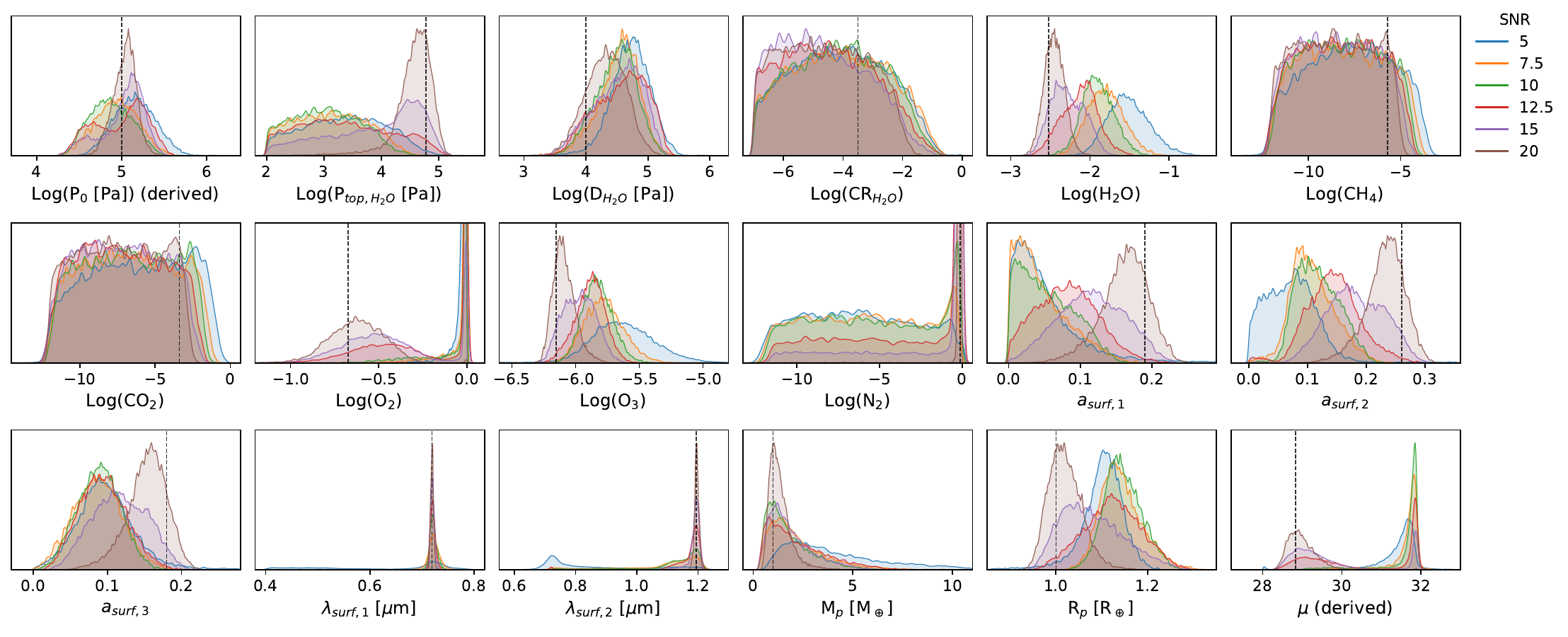}
    \caption{Full posterior distributions for the resolved-visible SNR sweep with $\RNIR = \num{70}$. Dashed lines indicate input values.}
    \label{fig:posteriors_SNR_R70_full}
\end{figure*}

Figure~\ref{fig:posteriors_SNR_R70_full} shows the full posterior distributions for all retrieved parameters and all SNR cases with $\RNIR = \num{70}$ discussed in Section~\ref{subsec:snr_resolved}.

\subsection{SNR sweep posteriors: resolved-visible ($\mathit{R_{NIR} = 40}$)}\label{app:snr_posteriors_r40}

\begin{figure*}[ht]
    \centering
    \includegraphics[width=\linewidth]{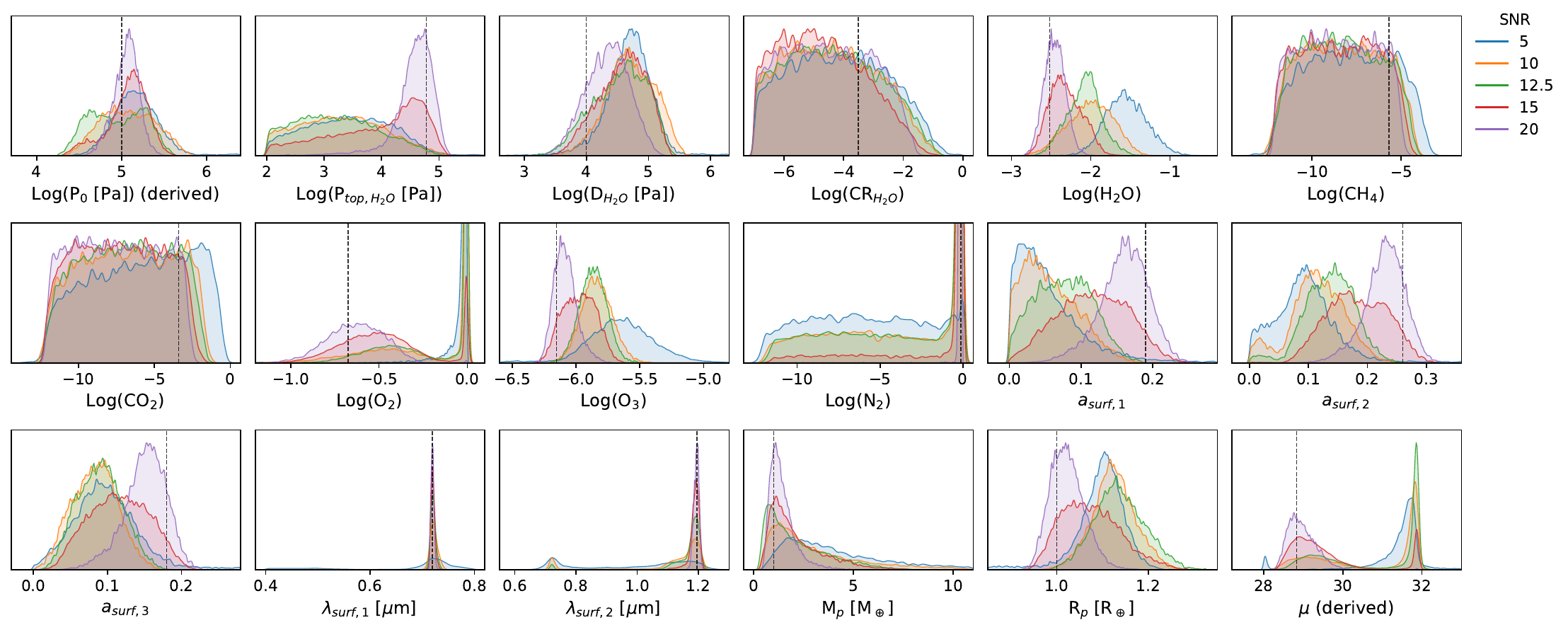}
    \caption{Posterior distributions for the resolved-visible SNR sweep with $\RNIR = \num{40}$. Dashed lines indicate input values.}
    \label{fig:posteriors_SNR_R40_full}
\end{figure*}

Figure~\ref{fig:posteriors_SNR_R40_full} shows the posterior distributions for the scenario with $\RNIR = \num{40}$ discussed in Section~\ref{subsec:snr_resolved}.

\subsection{High cloud-fraction posteriors (resolved-visible, $\mathit{R_{NIR} = 70}$)}\label{app:cloudfrac_posteriors}

\begin{figure*}[ht]
    \centering
    \includegraphics[width=\linewidth]{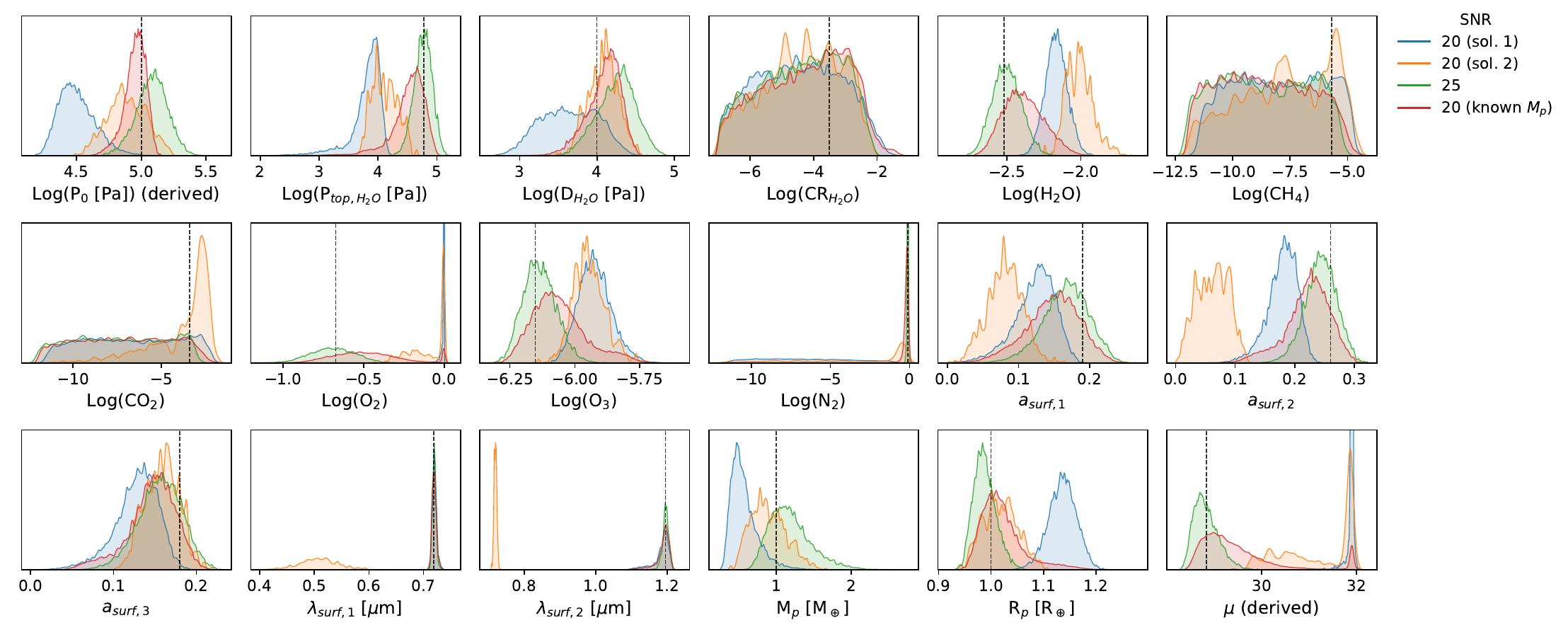}
    \caption{Posterior distributions for the \qty{50}{\percent} cloud-fraction scenario in the resolved-visible configuration. For $\mathrm{SNR} = \num{20}$ with free planetary mass, two solution modes are present. Dashed lines indicate input values.}
    \label{fig:posteriors_SNR_cld50}
\end{figure*}

Figure~\ref{fig:posteriors_SNR_cld50} shows the posterior distributions for the scenario with \qty{50}{\percent} cloud fraction and $\RNIR = \num{70}$ discussed in Section~\ref{subsec:snr_resolved}.

\subsection{SNR sweep posteriors: broadband-visible}\label{app:snr_posteriors_broadband}

\begin{figure*}[ht]
    \centering
    \includegraphics[width=\linewidth]{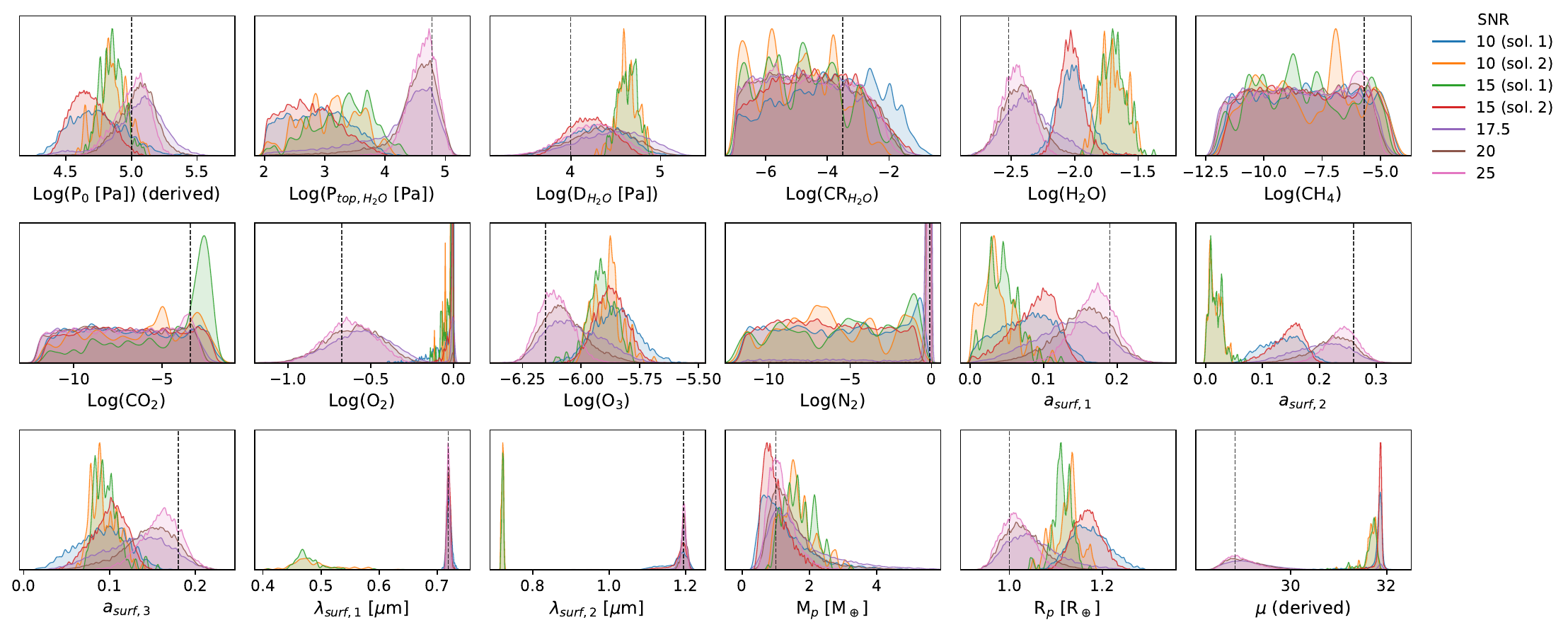}
    \caption{Full posterior distributions for the broadband-visible SNR sweep. For $\mathrm{SNR} = 10 \text{ and } \num{15}$, two solution modes are present. Dashed lines indicate input values.}
    \label{fig:posteriors_SNR_broadband_full}
\end{figure*}

Figure~\ref{fig:posteriors_SNR_broadband_full} shows the full posterior distributions for all retrieved parameters and all SNR cases discussed in Section~\ref{subsec:snr_broadband}.

\subsection{SNR sweep posteriors: broadband-visible with reduced near-IR coverage}\label{app:snr_posteriors_broadband_limited}

\begin{figure*}[ht]
    \centering
    \includegraphics[width=\linewidth]{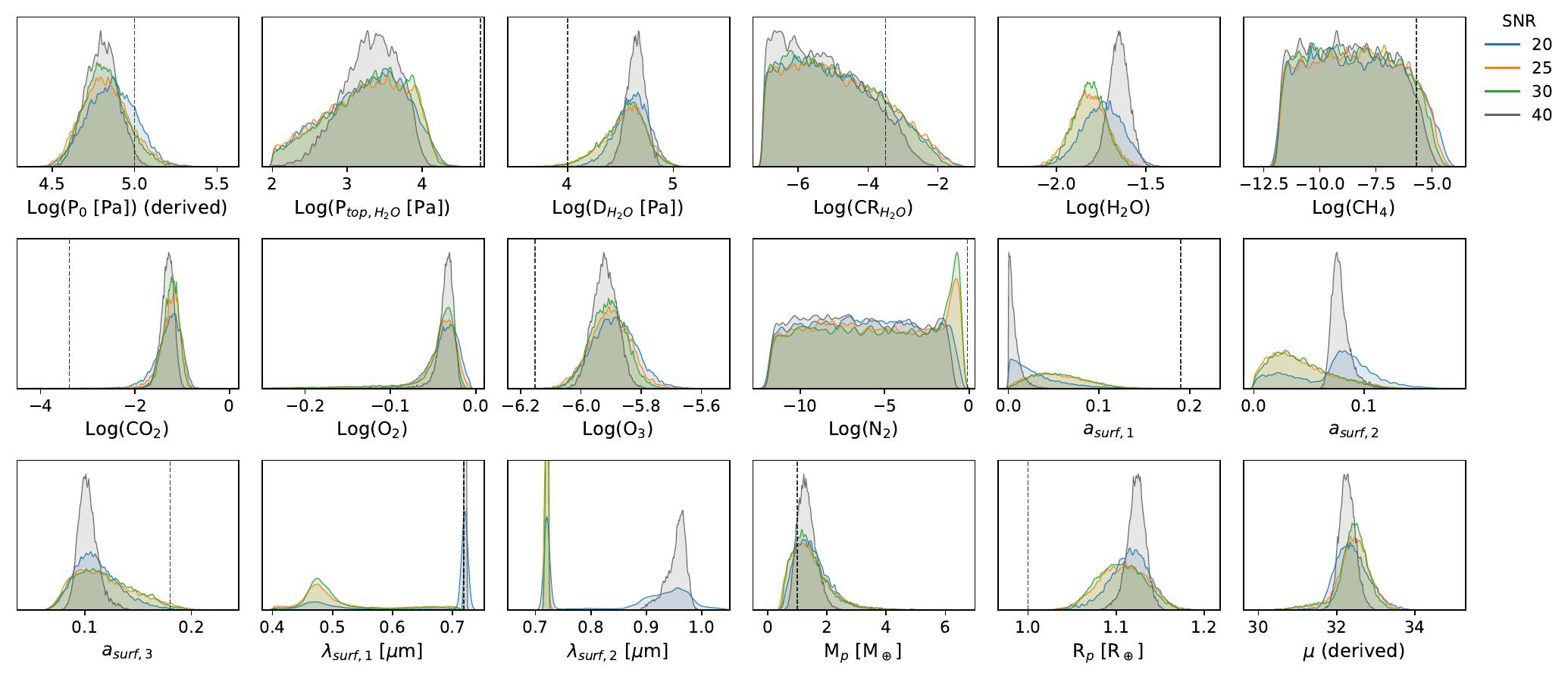}
    \caption{Full posterior distributions for the broadband-visible SNR sweep with reduced near-IR coverage (truncated at \qty{1.4}{\micro\meter}). Dashed lines indicate input values.}
    \label{fig:posteriors_SNR_broadband_limited_full}
\end{figure*}

Figure~\ref{fig:posteriors_SNR_broadband_limited_full} shows the full posterior distributions for all retrieved parameters and all SNR cases discussed in Section~\ref{subsubsec:snr_broadband_limited} for the broadband-visible configuration with the near-IR coverage truncated at \qty{1.4}{\micro\meter}.

\subsection{Radius-prior posteriors for reduced near-IR coverage}\label{app:radius_prior_posteriors}

\begin{figure*}[ht]
    \centering
    \includegraphics[width=\linewidth]{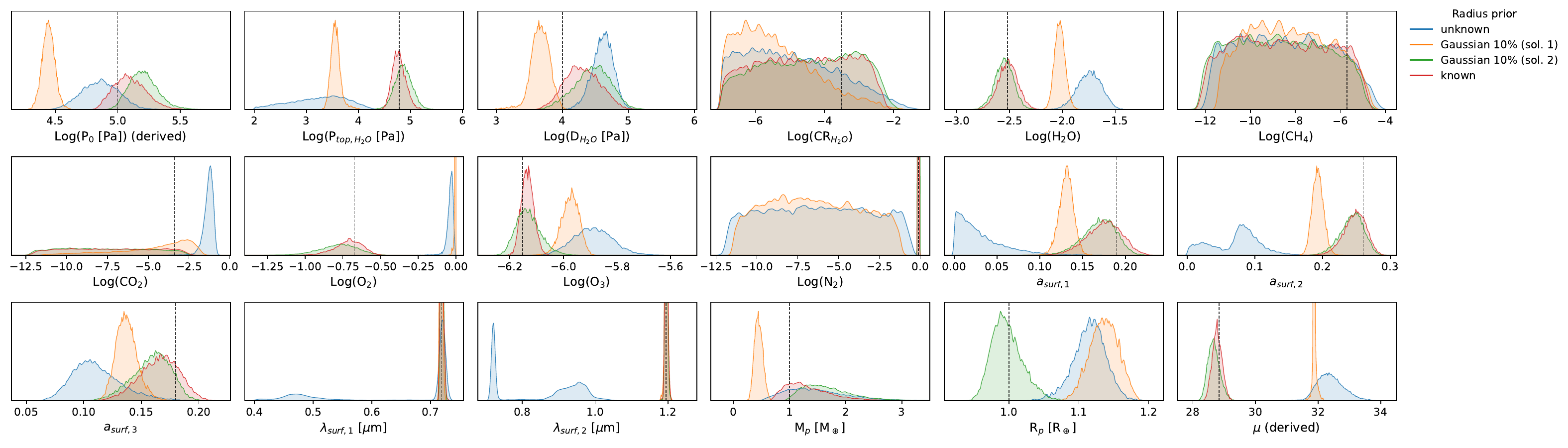}
    \caption{Posterior distributions for the broadband-visible configuration with reduced near-IR coverage (truncated at \qty{1.4}{\micro\meter}) for different radius assumptions. The cases shown are an unconstrained radius, a Gaussian radius prior centered on the input value with \qty{10}{\percent} standard deviation, and a fixed radius. For the Gaussian-prior case, two solution modes are shown separately. Dashed lines indicate input values.}
    \label{fig:posteriors_radius_prior}
\end{figure*}

Figure~\ref{fig:posteriors_radius_prior} shows the posterior distributions for supplemental reduced-coverage retrievals with an unconstrained radius, a Gaussian \qty{10}{\percent} radius prior, and fixed radius, as discussed in Section~\ref{subsec:discussion_obs}.

\bibliography{sample701}{}
\bibliographystyle{aasjournalv7}

\end{document}